# An Entropy Stable Formulation of Two-equation Turbulence Models with Particular Reference to the $k$-epsilon Model

Guillermo Hauke

I3A – Escuela de Ingenieria y Arquitectura
Universidad de Zaragoza
C/María de Luna 3
50018 Zaragoza, Spain

Thomas J. R. Hughes

Institute for Computational Engineering and Sciences
The University of Texas at Austin
201 East 24th Street, C0200
Austin, TX 78712, USA

April 13, 2025

## Abstract

Consistency and stability are two essential ingredients in the design of numerical algorithms for partial differential equations. Robust algorithms can be developed by incorporating nonlinear physical stability principles in their design, such as the entropy production inequality (i.e., the Clausius-Duhem inequality or second law of thermodynamics), rather than by simply adding artificial viscosity (a common approach). This idea is applied to the $k$-$\epsilon$ and two-equation turbulence models by introducing *space-time averaging.* Then, a set of entropy variables can be defined which leads to a symmetric system of advective-diffusive equations. Positivity and symmetry of the equations require certain constraints on the turbulence diffusivity coefficients and the turbulence source terms. With these, we are able to design entropy producing two-equation turbulence models and, in particular, the $k$-$\epsilon$ model.

## Contents

## 1. Introduction

In physics there exists a thermodynamic *stability principle* which governs the motion of fluids. This principle is the second law of thermodynamics, which postulates that the entropy of a closed system can only increase. The increase of entropy is due to irreversibilities of the system, such as viscous dissipation or heat transfer, caused by mechanical or thermal non-equilibrium states. Thus, entropy remains constant only if the process is reversible.

The principle of entropy production can be built into numerical schemes, allowing the physical stability principle of the system to be inherited by the numerical formulation. When using the Navier-Stokes equations to solve the fluid motion, embedding the entropy production principle into the formulation can be naturally achieved by using a variational formulation based on the so-called *entropy variables* [1,2,3,4,5].

However, most flows of interest are turbulent, in which case, although the Navier-Stokes equations still model the behavior of such flows, solution at all scale becomes out of reach of today's computers or simply impractical for design purposes. An often followed alternative is to average the equations in time, giving rise to the Reynolds-averaged Navier-Stokes equations (RANS).

Unfortunately, the meaning and the utility of the concept of entropy is obscure when we have to deal with the Reynolds-averaged equations [6] since, as we will see, the connection between the averaged entropy and the average of the dependent variables is lost. Nevertheless, interest in the entropy production characteristics of RANS models has been increasing, see, e.g., [8– 19]. Earlier, Jansen *et al.* [20,21,22] introduced a modified averaged entropy function with the corresponding entropy variables which returns the stability principle to the system and therefore can be built subsequently into the discrete formulation. This generalized entropy function is only useful when an equation for the square root of the turbulent kinetic energy or six equations for the square root of the Reynolds stress tensor are appended to the Reynolds-averaged Navier-Stokes equations.

However, it is more typical in practice to solve the Reynolds-averaged Navier-Stokes equations with an equation for the dissipation rate of turbulent kinetic energy [6,7]. In this case, the system of equations does not fit into the prior entropy producing format nor does there exist a suitable generalized entropy function. If the system of equations has to reproduce the physical principle of entropy production (which still holds for the averaged fluid motion) then there must be a generalized entropy function and the corresponding entropy variables which build the stability principle into the equations.

In this paper we develop these ideas for two-equation turbulence models and apply them specifically to the $k$-$\epsilon$ model. In Section 2, we review the basic concepts on symmetrization of nonlinear advective-diffusive systems. In Section 3, we introduce the *space-time averaging operator*, which leads to a symmetric formulation of the $k$-$\epsilon$ turbulence model (Section 4) and the corresponding generalized entropy function. In Section 5, an entropy production principle is obtained for standard assumptions of the $k$-$\epsilon$ model. This result is applied to the high-Reynolds number $k$-$\epsilon$ model in Sections 6, 7 and 8, and conclusions are drawn in Section 9.

## 2. Generalized entropy functions

Consider a hyperbolic system of conservation laws

$$\boldsymbol{U}_{,t} + \boldsymbol{F}^{\text{adv}}_{i,i} = \boldsymbol{0} \tag{1}$$

where $\boldsymbol{U}$ are the conservation variables (i.e., the conserved quantities) and $\boldsymbol{F}^{\text{adv}}_{i}$ is the advective flux in the $i^{\text{th}}$ direction. In quasi-linear form (1) can be expressed as

$$\boldsymbol{U}_{,t} + \boldsymbol{\mathcal{A}}_i \boldsymbol{U}_{,i} = \boldsymbol{0} \tag{2}$$

with $\boldsymbol{\mathcal{A}}_i = \boldsymbol{F}^{\text{adv}}_{i,\boldsymbol{U}}$ the Euler Jacobian with respect to conservation variables.

A *generalized entropy function* $\mathcal{H}(\boldsymbol{U})$ is a scalar-valued function satisfying these two properties:

*i.* $\mathcal{H}(\boldsymbol{U})$ is a strictly convex function of the conservation variables.

*ii.* There exists scalar-valued functions $\sigma_i = \sigma_i(\boldsymbol{U})$ called the entropy fluxes such that

$$\mathcal{H}_{,\boldsymbol{U}} \boldsymbol{\mathcal{A}}_i = \sigma_{i,\boldsymbol{U}}$$

Generalized entropy functions are related to an additional conservation law, an *entropy inequality*, which derives from the system of equations. Multiplying (1) by $\mathcal{H}_{,\boldsymbol{U}}$ on the left,

$$\mathcal{H}_{,\boldsymbol{U}}(\boldsymbol{U}_{,t} + \boldsymbol{F}^{\text{adv}}_{i,i}) = 0 \tag{3}$$

yields for weak solutions

$$\boxed{\mathcal{H}_{,t} + \sigma_{i,i} \leq 0} \tag{4}$$

which after integration in the space-time domain results in the *Clausius-Duhem inequality.* Note that weak solutions may include discontinuities, which are not differentiable. Thus, the equal sign applies only to smooth solutions.

There is a close connection between the existence of generalized entropy functions and *symmetric* hyperbolic systems. The following two theorems, presented in Tadmor [23], relate the two concepts.

**Theorem** (Mock [24])

*A hyperbolic system of conservation laws possessing a generalized entropy function $\mathcal{H}(\boldsymbol{U})$ becomes symmetric hyperbolic under the change of variables*

$$\boldsymbol{V}^T = \frac{\partial \mathcal{H}}{\partial \boldsymbol{U}} \tag{5}$$

**Theorem** (Godunov [25])

*If a hyperbolic system can be symmetrized by introducing a change of variables, then a generalized entropy function and corresponding entropy fluxes exist for this system.*

Thus, if a generalized entropy function $\mathcal{H}(\boldsymbol{U})$ exists, under the change of variables $\boldsymbol{U} \mapsto \boldsymbol{V}(\boldsymbol{U})$ the quasi-linear form (2) becomes a *symmetric* hyperbolic system

$$\boldsymbol{A}_0 \boldsymbol{V}_{,t} + \boldsymbol{A}_i \boldsymbol{V}_{,i} = \boldsymbol{0} \tag{6}$$

where

$$\begin{aligned} \boldsymbol{A}_0 &= \boldsymbol{U}_{,\boldsymbol{V}} \\ \boldsymbol{A}_i &= \boldsymbol{\mathcal{A}}_i \boldsymbol{A}_0 \end{aligned} \tag{7}$$

That is, the coefficient matrices enjoy the special properties:

*i.* $\boldsymbol{A}_0$ is symmetric positive-definite.

*ii.* The $\boldsymbol{A}_i$'s are symmetric.

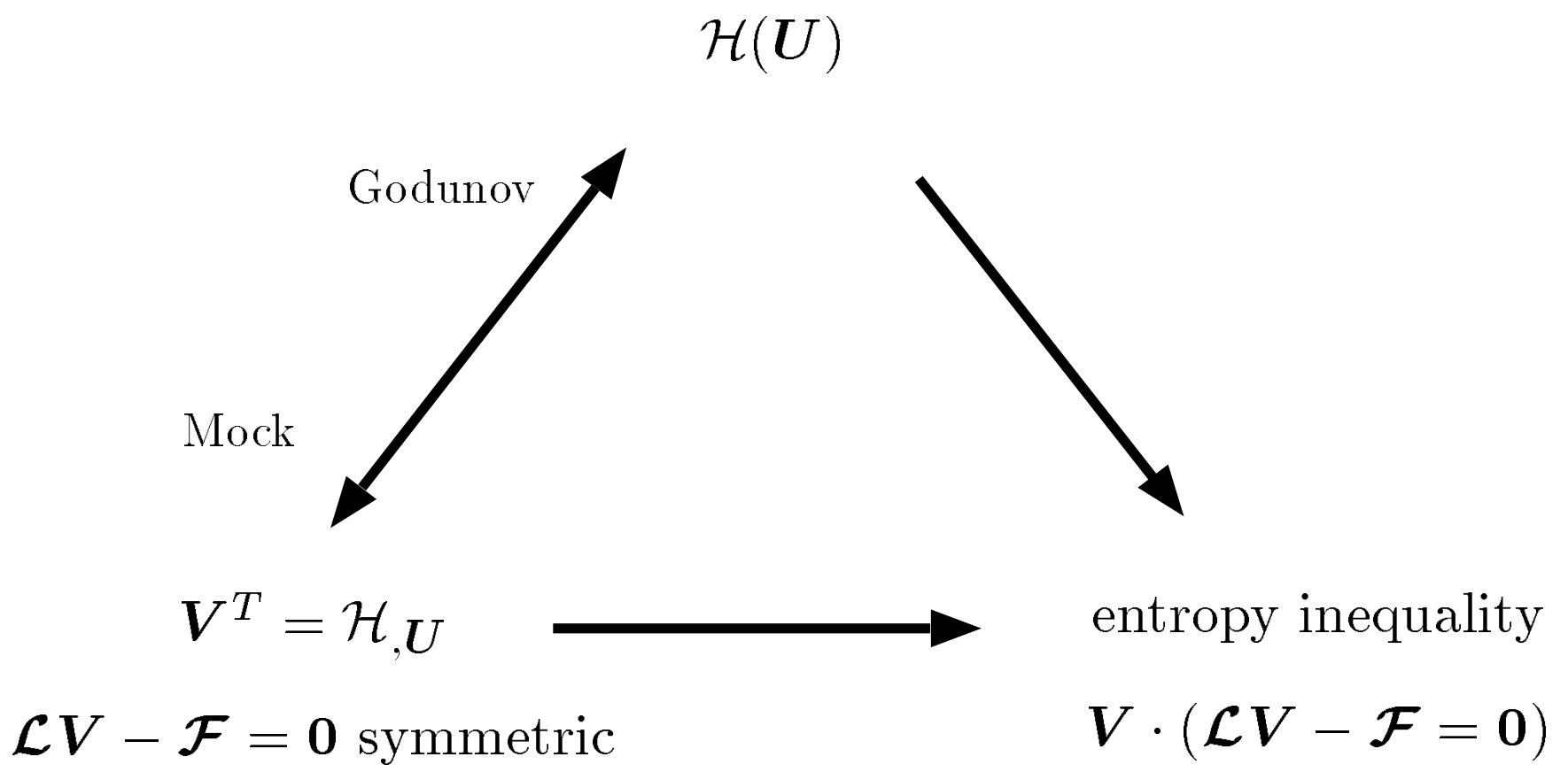

*Figure 1.* Relation among generalized entropy functions, entropy inequalities and symmetry of the quasi-linear form.

Figure 1 summarizes the relation between the three concepts discussed in this section: generalized entropy functions, entropy inequalities and symmetric advective diffusive systems. So far, we have neglected diffusive mechanisms. In the presence of diffusion, the symmetrization process involves one more term, whose effect will be discussed in the following subsection. Next, the symmetrization process for the Reynolds-averaged Navier-Stokes equations with a one-equation turbulence model will be reviewed, prior to extending the idea to two-equation models.

## 2.1. Navier-Stokes equations

The Navier-Stokes equations are obtained by addition of the diffusive fluxes and possibly a source term to the hyperbolic system (1)

$$\boldsymbol{U}_{,t} + \boldsymbol{F}^{\text{adv}}_{i,i} = \boldsymbol{F}^{\text{diff}}_{i,i} + \boldsymbol{\mathcal{S}} \tag{8}$$

Harten [26] determined a family of generalized entropy functions for (8) with viscosity terms included but heat conduction excluded. This work was extended by Hughes *et al.* [1] to include heat conduction, where it was shown that the *only* generalized entropy function was the *physical entropy* up to an affine transformation,

$$\mathcal{H}(\boldsymbol{U}) = -\rho s \tag{9}$$

Recall that the vector of conservation variables $\boldsymbol{U}$ for the compressible Navier-Stokes equations is

$$\boldsymbol{U} = \rho \begin{Bmatrix} 1 \\ u_1 \\ u_2 \\ u_3 \\ e_{\text{tot}} \end{Bmatrix} \tag{10}$$

where $\rho$ is the density, $u_i$ are the cartesian velocity components and $e_{\text{tot}}$ is the total energy per unit mass, which is the sum of the internal energy $e$ and the kinetic energy:

$$e_{\text{tot}} = e + \frac{1}{2} u_i u_i$$

For a general divariant gas, Chalot *et al.* [2] showed that the entropy variables

$\boldsymbol{V}^T = \mathcal{H}(\boldsymbol{U})_{,\boldsymbol{U}}$ are

$$\boldsymbol{V} = \frac{1}{T}\begin{Bmatrix} \mu - |\boldsymbol{u}|^2/2 \\ u_1 \\ u_2 \\ u_3 \\ -1 \end{Bmatrix} \tag{11}$$

where $\mu = e + p/\rho - Ts$ is the electrochemical potential per unit mass, $p$ is the pressure, $s$ is the entropy and $T$ is the absolute temperature. Multiplying (8) by $\mathcal{H}_{,\boldsymbol{U}}$, i.e., taking the dot product of the system with the entropy variables, results after some algebra in

$$\boxed{\begin{gathered}(\rho s)_{,t} + (\rho s u_i)_{,i} + \left(\frac{-\kappa T_{,i}}{T}\right)_{,i} = \left(\frac{\Upsilon(\boldsymbol{u},\boldsymbol{u})}{T}\right) + \kappa\frac{T_{,i}T_{,i}}{T^2} \\ \geq 0\end{gathered}} \tag{12}$$

where $\kappa$ is the heat conductivity coefficient and

$$\Upsilon(\boldsymbol{u},\boldsymbol{u}) = 2\mu^{\text{visc}} S^d_{ij}(\boldsymbol{u}) S^d_{ij}(\boldsymbol{u}) + \mu^{\text{visc}}_b S_{ll}(\boldsymbol{u}) S_{ll}(\boldsymbol{u}) \tag{13}$$

is the dissipation function, both positive-definite quantities. The expression has been written as a function of the strain rate tensor and its deviatoric part,

$$\begin{aligned} S_{ij}(\boldsymbol{u}) &= \frac{1}{2}(u_{i,j} + u_{j,i}) \\ S^d_{ij}(\boldsymbol{u}) &= S_{ij}(\boldsymbol{u}) - \frac{1}{3}S_{ll}(\boldsymbol{u})\delta_{ij} \end{aligned} \tag{14}$$

In obtaining the Clausius-Duhem inequality, the Gibbs relation

$$Tds = de + p\, d\frac{1}{\rho} \tag{15}$$

plays a fundamental role.

The set of entropy variables not only are the integration factors to recover the Clausius-Duhem inequality from the Navier-Stokes equations but also symmetrize the quasi-linear advective-diffusive operator. Under the change of variables $\boldsymbol{U} \mapsto \boldsymbol{V}(\boldsymbol{U})$, the quasi-linear form of the Navier-Stokes equations becomes a *symmetric* advective-diffusive system

$$\boldsymbol{A}_0 \boldsymbol{V}_{,t} + \boldsymbol{A}_i \boldsymbol{V}_{,i} = (\boldsymbol{K}_{ij}\boldsymbol{V}_{,j})_{,i} + \boldsymbol{\mathcal{S}} \tag{16}$$

where the coefficient matrices for entropy variables can be calculated from those for conservation variables, i.e., $\boldsymbol{\mathcal{A}}_i$ (Euler Jacobians) and $\boldsymbol{\mathcal{K}}_{ij}$ (diffusive matrices such that $\boldsymbol{F}_i^{\text{diff}} = \boldsymbol{\mathcal{K}}_{ij}\boldsymbol{U}_{,j}$),

$$\begin{aligned} \boldsymbol{A}_0 &= \boldsymbol{U}_{,\boldsymbol{V}} \\ \boldsymbol{A}_i &= \boldsymbol{\mathcal{A}}_i \boldsymbol{A}_0 \\ \boldsymbol{K}_{ij} &= \boldsymbol{\mathcal{K}}_{ij} \boldsymbol{A}_0 \end{aligned} \tag{17}$$

which enjoy the special properties:

*i.* $\boldsymbol{A}_0$ is symmetric positive-definite.

*ii.* The $\boldsymbol{A}_i$'s are symmetric.

*iii.* $\boldsymbol{K} = [\boldsymbol{K}_{ij}]$ is symmetric positive-semidefinite.

### 2.2. Reynolds-averaging with a one-equation model

Many flows of interest are turbulent and, although the instantaneous Navier-Stokes equations provide the full description of such flows, the range of scales is so wide that numerical computations become prohibitive. In order to reduce the size of the problem, it is common to average the equations, for example by means of ensemble or time averaging, giving rise to the Reynolds-averaged Navier-Stokes equations (RANS). The time averaging operator can be defined as

$$\overline{f}(\boldsymbol{x}, t) = \lim_{\mathcal{T}\to\infty} \frac{1}{\mathcal{T}} \int_{t-\mathcal{T}/2}^{t+\mathcal{T}/2} f(\boldsymbol{x}, t')dt' \tag{18}$$

For practical purposes, the integration time scale $\mathcal{T}$ can be taken such that it is smaller than variations of mean quantities but larger than several times the longest cycle of turbulence, so that (18) captures meaningful mean values. The compressible Reynolds-averaged equations are greatly simplified if they are expressed as a function of the *mass weighted average* or *Favre average* [28], defined as

$$\tilde{f}(\boldsymbol{x}, t) = \lim_{\mathcal{T}\to\infty} \frac{1}{\overline{\rho}\mathcal{T}} \int_{t-\mathcal{T}/2}^{t+\mathcal{T}/2} \rho(\boldsymbol{x}, t') f(\boldsymbol{x}, t')dt' \tag{19}$$

Both types of averages give rise to their respective fluctuations,

$$\begin{aligned} f &= \overline{f} + f' \\ f &= \tilde{f} + f'' \end{aligned} \tag{20}$$

One of the lowest levels of turbulence modeling appends one equation, say for the turbulent kinetic energy $k$, to the Reynolds-averaged Navier-Stokes equations. In this case, the vector of conserved quantities may become

$$\boldsymbol{U} = \overline{\rho} \left\{ \begin{array}{c} 1 \\ \tilde{u}_1 \\ \tilde{u}_2 \\ \tilde{u}_3 \\ \tilde{e}_{\text{tot}} \\ k \end{array} \right\} \tag{21}$$

where the Favre averaged total energy has contributions from the mean kinetic energy and the turbulent kinetic energy $k$, i.e.,

$$\tilde{e}_{\text{tot}} = \tilde{e} + \frac{1}{2}(\tilde{u}_i\tilde{u}_i + k) \tag{22}$$

Undoubtly, the second law of thermodynamics still applies to physical systems undergoing turbulence. However, as pointed out by Lesieur [6], the usefulness of the entropy production principle is not clear because the direct link between the averaged entropy $\tilde{s}$ and the Reynolds-averaged variables $\overline{\rho}$, $\tilde{u}_i$, $\widetilde{T}$, $\overline{p}$ is lost. Since entropy is defined by a nonlinear expression, namely the Gibbs relation,

$$T ds = de + p \; d\frac{1}{\rho} \tag{23}$$

its time average involves not only averaged variables but correlations between fluctuations as well. Thus, it is fruitless to find an exact Clausius-Duhem inequality for the averaged entropy, $\tilde{s}$, which relates only *Reynolds-averaged variables.*

But if entropy production has to play a fundamental role in the stability of turbulent flows, we may assume that some kind of stability estimate has to govern the system of equations for the *averaged variables* and therefore, an appropriate concept of generalized entropy must exist. Equivalently, the problem reduces to finding integration factors, that is, entropy variables, such that when multiplied by the Reynolds-averaged equations, an entropy inequality will follow and, at the same time, symmetrize the quasi-linear form of the system of equations.

Jansen *et al.* [20] realized that by changing variables from $k$ to a turbulent velocity scale

$$q_0 = \sqrt{2k} \tag{24}$$

the averaged total energy becomes

$$\tilde{e}_{\text{tot}} = \tilde{e} + \frac{1}{2}(\tilde{u}_i \tilde{u}_i + q_0 q_0) \tag{25}$$

With this change of variables, the turbulent velocity scale $q_0$ plays the same role as the physical velocity scales $u_i$. Defining the conservation variables as

$$\boldsymbol{U} = \overline{\rho} \left\{ \begin{array}{c} 1 \\ \tilde{u}_1 \\ \tilde{u}_2 \\ \tilde{u}_3 \\ \tilde{e}_{\text{tot}} \\ q_0 \end{array} \right\} \tag{26}$$

it is observed that they are the time average of the instantaneous conservation variables $\boldsymbol{U}$ in (10), where $q_0$ is just another velocity scale. Thus, defining the generalized entropy function as before, but as a function of averaged variables,

$$\mathcal{H}(\boldsymbol{U}) = -\overline{\rho}\hat{s} \tag{27}$$

where the following Gibbs-like relation is employed to compute $\hat{s}$

$$\widetilde{T} d\hat{s} = d\tilde{e} + \overline{p}\; d\frac{1}{\overline{\rho}} \tag{28}$$

and the modified electrochemical potential is also a function of averaged quantities:

$$\hat{\mu} = \tilde{e} + \overline{p}/\overline{\rho} - \widetilde{T}\hat{s} \tag{29}$$

gives rise to the entropy variables

$$\boldsymbol{V} = \frac{1}{\widetilde{T}} \left\{ \begin{array}{c} \hat{\mu} - |\tilde{\boldsymbol{u}}|^2/2 - q_0^2/2 \\ \tilde{u}_1 \\ \tilde{u}_2 \\ \tilde{u}_3 \\ -1 \\ q_0 \end{array} \right\} \tag{30}$$

which symmetrizes the system of equations.

Taking the dot product of the entropy variables with the system of equations results under the modeling assumptions of [20] in

$$\boxed{(\overline{\rho}\hat{s})_{,t} + (\overline{\rho}\hat{s}\tilde{u}_i)_{,i} + \left(\frac{-(\kappa+\kappa_T)\widetilde{T}_{,i}}{\widetilde{T}}\right)_{,i} = \left(\frac{\Upsilon(\tilde{\boldsymbol{u}},\tilde{\boldsymbol{u}})}{\widetilde{T}}\right) + (\kappa+\kappa_T)\,\frac{\widetilde{T}_{,i}\widetilde{T}_{,i}}{\widetilde{T}^2} + \left(\frac{\overline{\rho}\epsilon}{\widetilde{T}}\right) \geq 0} \tag{31}$$

Here $\kappa_T$ is the turbulent thermal conductivity and $\epsilon$ the dissipation rate of turbulent kinetic energy.

The above result has been obtained in [27] from a different point of view, namely, a formal analogy with the kinetic theory or rarefied gases.

Compared to the laminar counterpart (12), (31) indicates that turbulence results in an increased production of entropy due to the dissipation rate of turbulent kinetic energy $\epsilon$ and to the added turbulent thermal conductivity.

The generalized entropy $\hat{s}$ depends only on averaged variables. The usefulness of the concept of entropy is recovered, and conclusions concerning the impact of turbulence modeling assumptions on the production of the generalized entropy $\hat{s}$ can be drawn. Fur further discussion, see [6,20].

### 2.3. Reynolds-averaging with a two-equation model: $k$-$\epsilon$

One-equation models are appropriate for flows where the length scale is simple and can be prescribed algebraically. However, for complex flows such as recirculating flows with massive separation and three-dimensional flows, it becomes rather difficult to compute the algebraic length scale. Introducing a second partial differential equation for another turbulent variable can facilitate the computation of the length scale, which can be obtained as a function of the two turbulent quantities. The dissipation rate of turbulent kinetic energy $\epsilon$ has been a preferred choice in this regard. In the same way as an entropy producing system has been developed for a one-equation model, it is appealing to extend the idea of entropy stability to two-equation models. This stability principle can be used to design turbulence models that are in some sense entropy producing and give rise to stable numerical schemes.

If an equation for the dissipation rate of turbulence kinetic energy is added to the system of equations, the conserved quantities may become

$$\boldsymbol{U} = \overline{\rho} \begin{Bmatrix} 1 \\ \tilde{u}_1 \\ \tilde{u}_2 \\ \tilde{u}_3 \\ \tilde{e}_{\text{tot}} \\ q_0 \\ \epsilon \end{Bmatrix} \tag{32}$$

The system of equations can be written as

$$\boldsymbol{\mathcal{L}U} - \boldsymbol{\mathcal{S}} = \boldsymbol{0} \tag{33}$$

We want to find the integration factors $\boldsymbol{V}$ such that the product

$$\boldsymbol{V} \cdot (\boldsymbol{\mathcal{L}U} - \boldsymbol{\mathcal{S}}) = \boldsymbol{0} \tag{34}$$

yields an entropy stability estimate.

For a general divariant gas, every thermodynamic property can be expressed as a function of two independent thermodynamic variables. Consequently, the generalized entropy function for the Reynolds-averaged Navier-Stokes equations can be expressed as a function of averaged density and temperature, $\hat{s} = \hat{s}(\overline{\rho}, \widetilde{T})$, or averaged density and internal energy $\hat{s} = \hat{s}(\overline{\rho}, \widetilde{e})$. The convexity of entropy with respect to the conserved quantities $\boldsymbol{U}$ is achieved by the dependency of the internal energy on the total energy and at the same time, the dependency of the total energy on the velocity components, i.e.,

$$\hat{s}(\boldsymbol{U}) = \hat{s}(\overline{\rho}, \tilde{e}) = \hat{s}\left(\overline{\rho}, [\tilde{e}_{\text{tot}} - \frac{1}{2}\left(|\tilde{\boldsymbol{u}}|^2 + q_0^2\right)]\right) \tag{35}$$

The entropy variables are computed from $\hat{s}$ by taking the partial derivatives

$$\boldsymbol{V}^T = \frac{\partial \mathcal{H}(\boldsymbol{U})}{\partial \boldsymbol{U}} \tag{36}$$

or in components

$$V_i = \mathcal{H}(\boldsymbol{U})_{,U_i} \tag{37}$$

In this way, velocity scales and entropy become linked. However, in this framework, $\epsilon$ is playing no role since $\hat{s}$ and $\tilde{e}_{\text{tot}}$ do not depend on $\epsilon$ and so the corresponding entropy variable is zero,

$$\begin{aligned} V_7 &= \mathcal{H}(\boldsymbol{U})_{,U_7} \\ &= (\overline{\rho}\hat{s})_{,(\overline{\rho}\epsilon)} \\ &= 0 \end{aligned} \tag{38}$$

Hence, a link between $\epsilon$ and the generalized entropy function needs to be developed.

## 3. Space-time averaging

Reynolds-averaging takes into account the effect of temporal averaging into the mean flow. However, turbulence is characterized by temporal and spatial fluctuations. Thus, it is natural to think of the averaging process as including both, temporal and spatial averaging. In fact, space-averaging of the turbulent kinetic energy provides the link between the averaged total energy and the dissipation rate of turbulent kinetic energy, $\epsilon$.

### 3.1. Space averaged turbulent kinetic energy

In the previous section, it was concluded that, in order to generate a set of entropy variables, a link between the generalized entropy function and $\epsilon$ needs to be developed. In this section we show that, in fact, $\epsilon$ is accounting for spatial averaging of the turbulent kinetic energy $k$. Let us calculate the spatial average of the turbulent kinetic energy $\langle k\rangle$ over a small spatial domain, Vol, centered about $(\bar{\boldsymbol{x}}, t)$,

$$2\langle k\rangle(\bar{\boldsymbol{x}}, t) = \lim_{\mathcal{T}\to\infty} \frac{1}{\overline{\rho}\mathcal{T}\text{Vol}} \int_{t-\mathcal{T}/2}^{t+\mathcal{T}/2} \int_{\text{Vol}} \rho(\boldsymbol{x},\tau) u_i''(\boldsymbol{x},\tau) u_i''(\boldsymbol{x},\tau)\; d\text{Vol}d\tau \tag{39}$$

Expanding $u_i''$ in a Taylor series about $\bar{\boldsymbol{x}}$,

$$u_i''(\boldsymbol{x}, t) = u_i''(\bar{\boldsymbol{x}}, t) + (x_j - \bar{x}_j)\frac{\partial u_i''}{\partial x_j}(\bar{\boldsymbol{x}}, t) + O((x_j - \bar{x}_j)^2) \tag{40}$$

and carrying out the integration under the assumption of slowly varying density within Vol yields

$$2\langle k\rangle(\bar{\boldsymbol{x}},t) = \widetilde{u_i''(\bar{\boldsymbol{x}},t)u_i''}(\bar{\boldsymbol{x}},t) + I_{jk}\widetilde{\frac{\partial u_i''}{\partial x_j}(\bar{\boldsymbol{x}},t)\frac{\partial u_i''}{\partial x_k}}(\bar{\boldsymbol{x}},t) \tag{41}$$

where $I_{jk}$ is the inertia tensor of the volume

$$I_{jk} = \frac{1}{\mathrm{Vol}}\int_{\mathrm{Vol}} (x_j - \bar{x}_j)(x_k - \bar{x}_k)\ d\mathrm{Vol} \tag{42}$$

If the volume of spatial averaging is a cube of side $h$, then

$$I_{jk} = \frac{1}{12}h^2\delta_{jk} \tag{43}$$

and therefore

$$2\langle k\rangle(\bar{\boldsymbol{x}},t) = \widetilde{u_i''u_i''}(\bar{\boldsymbol{x}},t) + \frac{1}{12}h^2\widetilde{\frac{\partial u_i''}{\partial x_j}\frac{\partial u_i''}{\partial x_j}}(\bar{\boldsymbol{x}},t) \tag{44}$$

Substituting in (44) the definition of $q_0$, the Favre-averaged turbulent kinetic energy,

$$q_0^2 = \widetilde{u_i''u_i''} \tag{45}$$

and the definition of $\epsilon$

$$\epsilon = \nu\widetilde{\frac{\partial u_i''}{\partial x_j}\frac{\partial u_i''}{\partial x_j}} \tag{46}$$

with $\nu$ the kinematic viscosity, results in

$$2\langle k\rangle(\bar{\boldsymbol{x}},t) = q_0^2 + \frac{1}{12}h^2\frac{\epsilon}{\nu} \tag{47}$$

Although the previous definition of $\epsilon$ is widely used, strictly speaking it corresponds to homogeneous, incompressible turbulence, with negligible viscosity-velocity gradient correlations (see Hinze [29]).

If we define a new velocity scale $q_1$ such that

$$q_1^2 = \frac{1}{12}h^2\frac{\epsilon}{\nu} \tag{48}$$

then the space average of the turbulent kinetic energy can be written as

$$2\langle k\rangle(\bar{\boldsymbol{x}},t) \stackrel{\mathrm{def}}{=} q_0^2 + q_1^2 \tag{49}$$

Still $q_1$ is not determined since it depends upon the choice of the volume length $h$, which has to be large enough to capture the small scales but small enough to be insensitive to variations of averaged quantities. If we choose $h$ proportional to the Kolmogorov length scale,

$$h = \alpha \left( \frac{\nu^3}{\epsilon} \right)^{\frac{1}{4}} \tag{50}$$

it turns out that for $\alpha = \sqrt{12}$

$$q_1 = (\nu \epsilon)^{\frac{1}{4}} \tag{51}$$

Note that the choice of $h$ is arbitrary, so different definitions will lead to different variables $q_1$. Dimensional analysis shows that if $q_1 = q_1(q_0, \epsilon, \nu)$ then $q_1$ is a function of the form

$$q_1 = q_0 f(\frac{\nu \epsilon}{q_0^4}) \tag{52}$$

As pointed out by Bradshaw [30], it is equally plausible to try the Taylor microscale

$$\begin{aligned} h^2 = \lambda^2 &= \frac{q_0^2/3}{\overline{\left( \frac{\partial u_1'}{\partial x_1} \right)^2}} \\ &= 15\nu \frac{q_0^2/3}{\epsilon} \end{aligned} \tag{53}$$

where the last equality holds for isotropic turbulence [29] (small-scale structure tends to be isotropic at large Reynolds number). Then $q_1$ turns out to be proportional to $q_0$,

$$q_1 = \sqrt{\frac{15}{12 \times 3}}\, q_0 \tag{54}$$

and we cannot extract more information from spatial averaging since we are reverting to the one-equation model case.

A similar decomposition is proposed by Wilcox [31], where the turbulent kinetic energy is split into the large energy-containing eddies contribution and the small eddies contribution. The small eddies are assumed to be isotropic and are quickly dissipated into heat, while the large eddies are supposed to behave almost inviscidly. Here, $q_0$ is the contribution to the turbulent kinetic energy due to time averaging and $q_1$ the contribution due to spatial averaging.

### 3.2. Space-time averaged Navier-Stokes equations

Let us redefine the Reynolds and Favre averaging operators as space-time averaging operators:

$$\overline{f}(\boldsymbol{x},t) = \lim_{\mathcal{T}\to\infty} \frac{1}{\mathcal{T}\mathrm{Vol}} \int_{t-\mathcal{T}/2}^{t+\mathcal{T}/2} \int_{\mathrm{Vol}} f(\boldsymbol{x}',t') d\mathrm{Vol} dt' \tag{55}$$

$$\tilde{f}(\boldsymbol{x},t) = \lim_{\mathcal{T}\to\infty} \frac{1}{\overline{\rho}\mathcal{T}\mathrm{Vol}} \int_{t-\mathcal{T}/2}^{t+\mathcal{T}/2} \int_{\mathrm{Vol}} \rho f(\boldsymbol{x}',t') d\mathrm{Vol} dt' \tag{56}$$

with the corresponding fluctuations

$$f = \overline{f} + f' \tag{57}$$

$$f = \tilde{f} + f'' \tag{58}$$

These averaging operators do not satisfy the simple rules that a well-behaved statistical mean operator should possess. However, since the averaging-volume is very small, it will be still assumed that

$$\overline{\overline{f}} = \overline{f} \qquad \overline{\overline{f}g} = \overline{f}\overline{g} \tag{59}$$

$$\overline{\tilde{f}} = \tilde{f} \qquad \overline{\overline{f}\tilde{g}} = \overline{f}\tilde{g} \tag{60}$$

These hypothesis are not essential if one introduces the generalized central moments defined by Germano in [32] and consistently modifies the assumptions outlined below. If these operators are applied to the Navier-Stokes equations, we obtain the typical RANS, only this time the tildes and bars denote space-time averages,

$$\begin{aligned}
\overline{\rho}_{,t} + [\overline{\rho}\tilde{u}_i]_{,i} &= 0 \\
[\overline{\rho}\tilde{u}_i]_{,t} + [\overline{\rho}\tilde{u}_i\tilde{u}_j]_{,j} &= \Big[-\overline{p}\delta_{ij} + \tau_{ij}(\tilde{\boldsymbol{u}}) - \overline{\rho}\widetilde{u_i''u_j''} + \tau_{ij}(\overline{\boldsymbol{u}''})\Big]_{,j} + \overline{\rho}\tilde{b}_i \\
[\overline{\rho}\tilde{e}_{\mathrm{tot}}]_{,t} + [\overline{\rho}\tilde{u}_i\tilde{e}_{\mathrm{tot}}]_{,i} &= \Big[-\overline{p}\tilde{u}_i + \tau_{ij}(\tilde{\boldsymbol{u}})\tilde{u}_j + \kappa\tilde{T}_{,i} - \tilde{u}_j\overline{\rho}\widetilde{u_j''u_i''} \\
&\quad - \overline{\rho}\widetilde{u_i''e''} - \overline{pu_i''} - \frac{1}{2}\overline{\rho}\widetilde{u_j''u_j''u_i''} + \overline{\tau_{ij}(\boldsymbol{u}'')u_j''} \\
&\quad + \tau_{ij}(\overline{\boldsymbol{u}''})\tilde{u}_j + \tau_{ij}(\tilde{\boldsymbol{u}})\overline{u_j''} + \kappa\overline{T_{,i}''}\Big]_{,i} \\
&\quad + \overline{\rho}(\tilde{b}_i\tilde{u}_i + \widetilde{u_i''b_i''} + \tilde{r})
\end{aligned} \tag{61}$$

Here, we are denoting cartesian velocity components by $u_i$, density by $\rho$, pressure by $p$, temperature by $T$, stress tensor components by $\tau_{ij}$, total energy per unit mass by $e_{\rm tot}$, body force per unit mass by $b_i$, thermal heat conductivity by $\kappa$, and the heat source per unit volume by $r$. Now the average of the total energy per unit mass is

$$\begin{aligned}\tilde{e}_{\rm tot} &= \tilde{e} + \frac{1}{2}(\tilde{u}_i\tilde{u}_i + \widetilde{u''_i u''_i}) \\ &= \tilde{e} + \frac{1}{2}\tilde{u}_i\tilde{u}_i + \langle k\rangle \\ &= \tilde{e} + \frac{1}{2}(\tilde{u}_i\tilde{u}_i + q_0q_0 + q_1q_1)\end{aligned} \tag{62}$$

where the last equality emanates from definition (49).

We may assume by averaging in space and neglecting variations of diffusivity coefficients in the volume Vol that the usual modeling assumptions still apply by replacing $k$ with $\langle k\rangle$. In particular let us assume the following.

*i.* The eddy-viscosity assumption takes the form

$$-\overline{\rho}\widetilde{u''_i u''_j} = 2\mu_T^{\rm visc} S^d_{ij}(\tilde{\boldsymbol{u}}) - \frac{2}{3}\overline{\rho}\langle k\rangle\delta_{ij} \tag{63}$$

The last term maintains the correct trace of the space-time Reynolds-stress tensor, $2\overline{\rho}\langle k\rangle$. The turbulent viscosity is defined as before:

$$\mu_T^{\rm visc} = C_\mu f_\mu \overline{\rho}\frac{k^2}{\epsilon} \tag{64}$$

where $C_\mu$ is a constant and $f_\mu$ is a damping function depending on turbulent Reynolds number.

*ii.* Fluctuating velocity-energy correlations are modeled via the gradient diffusion hypothesis

$$\begin{aligned}-\overline{\rho}\widetilde{u''_i e''} &= c_v\frac{\mu_T^{\rm visc}}{Pr_T}\widetilde{T}_{,i} \\ &= \kappa_T\widetilde{T}_{,i}\end{aligned} \tag{65}$$

Likewise, the velocity-body force correlation,

$$-\overline{\rho}\widetilde{u''_i b''_i} = \frac{\mu_T^{\rm visc}}{Pr_b}\tilde{b}_{,i} \tag{66}$$

*iii.* Turbulent transport of $k$ and pressure diffusion are jointly approximated by

$$-\frac{1}{2}\overline{\rho}\widetilde{u''_i u''_i u''_j} - \overline{p' u''_j} = \frac{\mu_T^{\rm visc}}{Pr_k}\langle k\rangle_{,j} \tag{67}$$

which is generalized to

$$-\frac{1}{2}\overline{\rho}\widetilde{u_i''u_i''u_j''} - \overline{p'u_j''} = \frac{1}{2}\left(\frac{\mu_T^{\text{visc}}}{Pr_k}(q_0^2)_{,j} + \frac{\mu_T^{\text{visc}}}{Pr_{q_1}}(q_1^2)_{,j}\right) \tag{68}$$

The assumption consistent with the $k$-$\epsilon$ model corresponds to $Pr_k = Pr_{q_1}$.

*iv.* The molecular diffusion term assumes the form

$$\overline{u_i''\tau_{ij}(\boldsymbol{u}'')} = \mu^{\text{visc}}\langle k\rangle_{,j} \tag{69}$$

*v.* The pressure dilatation term is ignored,

$$\overline{p'u_{i,i}''} = 0 \tag{70}$$

*vi.* Compressible terms, i.e., averages of single fluctuations, are neglected:

$$\tau_{ij}(\overline{\boldsymbol{u}''})\tilde{u}_j + \tau_{ij}(\tilde{\boldsymbol{u}})\overline{u_j''} + \kappa\overline{T_{,i}''} = 0 \tag{71}$$

*vii.* The dissipation rate of turbulent kinetic energy is defined by

$$\overline{\rho}\epsilon = \overline{\tau_{ij}(\boldsymbol{u}'')u_{i,j}''} \tag{72}$$

*vii.* We are also decomposing the spatial average of the turbulent kinetic energy as

$$\langle k\rangle = \frac{1}{2}(q_0^2 + q_1^2) \tag{73}$$

where

$$q_1 = (\nu\epsilon)^{\frac{1}{4}} \tag{74}$$

Note that $\langle k\rangle = \frac{1}{2}\widetilde{u_i''u_i''}$ is a space-time average and that $q_0$ and $q_1$ are time averages.

Substitution of these modeling assumptions in (61) results in the modeled space-time averaged Navier-Stokes equations,

$$\begin{aligned}
\overline{\rho}_{,t} + [\overline{\rho}\tilde{u}_i]_{,i} &= 0 \\
[\overline{\rho}\tilde{u}_i]_{,t} + [\overline{\rho}\tilde{u}_i\tilde{u}_j + \overline{p}\delta_{ij}]_{,j} &= \left[2(\mu^{\text{visc}} + \mu_T^{\text{visc}})S_{ij}^d(\tilde{\boldsymbol{u}}) + \mu_b^{\text{visc}}S_{ll}(\tilde{\boldsymbol{u}})\delta_{ij}\right]_{,j} \\
&\quad + \overline{\rho}\tilde{b}_i - [\frac{2}{3}\overline{\rho}\langle k\rangle]_{,i} \\
[\overline{\rho}\tilde{e}_{\text{tot}}]_{,t} + [\overline{\rho}\tilde{u}_i\tilde{e}_{\text{tot}} + \overline{p}\tilde{u}_i]_{,i} &= \Bigg[\tilde{u}_j\left(2(\mu^{\text{visc}} + \mu_T^{\text{visc}})S_{ij}^d(\tilde{\boldsymbol{u}}) + \mu_b^{\text{visc}}S_{ll}(\tilde{\boldsymbol{u}})\delta_{ij}\right) \\
&\quad + (\kappa + \kappa_T)\widetilde{T}_{,i} + \left(\mu^{\text{visc}} + \frac{\mu_T^{\text{visc}}}{\text{Pr}_k}\right)\langle k\rangle_{,i}\Bigg]_{,i} \\
&\quad + \overline{\rho}(\tilde{b}_i\tilde{u}_i - \frac{\mu_T^{\text{visc}}}{\text{Pr}_b}\tilde{b}_{i,i} + \tilde{r}) - [\frac{2}{3}\overline{\rho}\langle k\rangle\tilde{u}_i]_{,i}
\end{aligned} \tag{75}$$

where the total energy per unit mass is now

$$\tilde{e}_{\text{tot}} = \tilde{e} + \frac{1}{2}(\tilde{u}_i\tilde{u}_i + q_0q_0 + q_1q_1) \tag{76}$$

## 4. Symmetric advective-diffusive system for the $k$-$\epsilon$ model

In this section, we develop an entropy consistent $k$-$\epsilon$ model by introducing space-time averaging operators. The space-time averaged Navier-Stokes equations coupled with the $k$-$\epsilon$ model possess a generalized entropy function which defines a set of entropy variables and gives rise to a symmetric form of the system of equations.

The $k$-$\epsilon$ turbulence model appends two equations to the Reynolds-averaged Navier-Stokes equations to describe the effect of turbulence on the mean flow: an equation for the turbulent kinetic energy $k$ and another equation for the dissipation rate, $\epsilon$.

The equation for the turbulent kinetic energy can be expressed in a general form as

$$[\overline{\rho}k]_{,t} + [\overline{\rho}\tilde{u}_ik]_{,i} = [(\mu^{\text{visc}} + \frac{\mu_T^{\text{visc}}}{\text{Pr}_k})k_{,i}]_{,i} + \mathcal{S}_k \tag{77}$$

where the source term is given by

$$\mathcal{S}_k = 2\mu_T^{\text{visc}}S_{ij}^d(\tilde{\boldsymbol{u}})S_{ij}^d(\tilde{\boldsymbol{u}}) - \frac{2}{3}\overline{\rho}k\tilde{u}_{i,i} - \overline{\rho}\epsilon - \frac{\mu_T^{\text{visc}}}{\text{Pr}_b}\tilde{b}_{i,i} \tag{78}$$

To obtain an entropy-producing, symmetric system we change variables from $k$ to $q_0$ resulting in the equation [20]

$$[\overline{\rho}q_0]_{,t} + [\overline{\rho}\tilde{u}_iq_0]_{,i} = [(\mu^{\text{visc}} + \frac{\mu_T^{\text{visc}}}{\text{Pr}_k})q_{0,i}]_{,i} + \mathcal{S}_{q_0} \tag{79}$$

where

$$\begin{aligned}\mathcal{S}_{q_0} &= \frac{1}{q_0}(\mathcal{S}_k + (\mu^{\text{visc}} + \frac{\mu_T^{\text{visc}}}{\text{Pr}_k})q_{0,i}q_{0,i}) \\ &= \frac{1}{q_0}2\mu_T^{\text{visc}}S_{ij}^d(\tilde{\boldsymbol{u}})S_{ij}^d(\tilde{\boldsymbol{u}}) - \frac{1}{3}\overline{\rho}q_0\tilde{u}_{i,i} - \frac{1}{q_0}\overline{\rho}\epsilon \\ &\quad + \left(\mu^{\text{visc}} + \frac{\mu_T^{\text{visc}}}{\text{Pr}_k}\right)\frac{q_{0,i}q_{0,i}}{q_0} - \frac{\mu_T^{\text{visc}}}{q_0\,\text{Pr}_b}\tilde{b}_{i,i}\end{aligned} \tag{80}$$

Here, we are assuming that this equation is valid despite that the averaging operators are now space-time averages.

The modeled equation for $\epsilon$ can be cast in the general form

$$[\overline{\rho}\epsilon^*]_{,t} + [\overline{\rho}\tilde{u}_i\epsilon^*]_{,i} = [(\mu^{\text{visc}} + \frac{\mu_T^{\text{visc}}}{\text{Pr}_\epsilon})\epsilon^*_{,i}]_{,i} + \mathcal{S}_\epsilon \tag{81}$$

where

$$\epsilon = \epsilon^* + \epsilon_0 \tag{82}$$

$\epsilon^*$ represents the computed part of the dissipation and $\epsilon_0$ is a model-dependent function, which allows the setting of zero boundary conditions for $\epsilon^*$ at the wall. If $\epsilon_0 = 0$ then $\epsilon^*$ coincides with the true dissipation rate. The source term $\mathcal{S}_\epsilon$ includes the production $\mathcal{P}_\epsilon$, dissipation $\mathcal{D}_\epsilon$, and other possible terms such as near wall corrections and compressible terms $\mathcal{R}_\epsilon$,

$$\mathcal{S}_\epsilon = \mathcal{P}_\epsilon - \mathcal{D}_\epsilon + \mathcal{R}_\epsilon \tag{83}$$

For a general $k$-$\epsilon$ model these terms can be expressed as

$$\begin{aligned} \mathcal{P}_\epsilon &= C_{\epsilon 1} f_1 \frac{\epsilon}{k}\left(\mathcal{P}_k + \frac{2}{3}\overline{\rho}\tilde{u}_{i,i}k\right) \\ \mathcal{P}_k &= 2\mu_T^{\text{visc}} S^d_{ij}(\tilde{\boldsymbol{u}})S^d_{ij}(\tilde{\boldsymbol{u}}) \\ \mathcal{D}_\epsilon &= C_{\epsilon 2} f_2 \overline{\rho}\frac{\epsilon^*\epsilon^*}{k} \end{aligned} \tag{84}$$

where $C_{\epsilon 1}$, $C_{\epsilon 2}$ are model-dependent constants, $f_1$ and $f_2$ are model-dependent damping functions. Finally, the turbulent eddy viscosity is given by

$$\mu_T^{\text{visc}} = C_\mu f_\mu \overline{\rho}\frac{k^2}{\epsilon^*} \tag{85}$$

We also change variables from $\epsilon^*$ to $q_1$ such that

$$\epsilon^* \mapsto q_1 = (\nu\epsilon^*)^{1/4} \tag{86}$$

which yields

$$[\overline{\rho}q_1]_{,t} + [\overline{\rho}\tilde{u}_i q_1]_{,i} = [(\mu^{\text{visc}} + \frac{\mu_T^{\text{visc}}}{\text{Pr}_\epsilon})q_{1,i}]_{,i} + \mathcal{S}_{q_1} \tag{87}$$

where $\text{Pr}_\epsilon$ is a turbulent Prandtl number. The source term $\mathcal{S}_{q_1}$ absorbs all the non-advective-diffusive terms.

*Remarks*

1. Observe that all the variables, i.e. $\overline{\rho}$, $\tilde{u}_i$, $\overline{p}$, $\widetilde{T}$,..., are space-time averages except $q_0$ and $q_1$, which are only time averages. The variables $q_0$ and $q_1$ maintain their roles of standard Reynolds-averaging. Thus, the vector of unknows contains variables with two different types of averaging.

2. Other two-equation turbulence models can also be transformed into the $q_0$, $q_1$ equations using appropriate changes of variables, preserving the structure of the equations with different turbulence source terms $\mathcal{S}_{q_0}$, $\mathcal{S}_{q_1}$.

The system of equations can be written in the more compact form

$$\boldsymbol{U}_{,t} + \boldsymbol{F}^{\text{adv}}_{i,i} = \boldsymbol{F}^{\text{diff}}_{i,i} + \boldsymbol{\mathcal{S}} \tag{88}$$

All the vectors are made explicit below:

$$\boldsymbol{U} = \left\{ \begin{array}{c} U_1 \\ U_2 \\ U_3 \\ U_4 \\ U_5 \\ U_6 \\ U_7 \end{array} \right\} = \overline{\rho} \left\{ \begin{array}{c} 1 \\ \tilde{u}_1 \\ \tilde{u}_2 \\ \tilde{u}_3 \\ \tilde{e}_{\text{tot}} \\ q_0 \\ q_1 \end{array} \right\} \tag{89}$$

$$\boldsymbol{F}^{\text{adv}}_i = \tilde{u}_i \boldsymbol{U} + \overline{p} \left\{ \begin{array}{c} 0 \\ \delta_{1i} \\ \delta_{2i} \\ \delta_{3i} \\ \tilde{u}_i \\ 0 \\ 0 \end{array} \right\} \tag{90}$$

$$\boldsymbol{F}^{\text{diff}}_i = \left\{ \begin{array}{c} 0 \\ \tau_{1i} \\ \tau_{2i} \\ \tau_{3i} \\ \tau_{ij}\tilde{u}_j \\ 0 \\ 0 \end{array} \right\} + \left\{ \begin{array}{c} 0 \\ 0 \\ 0 \\ 0 \\ -q^{\text{heat}}_i \\ 0 \\ 0 \end{array} \right\} + \left\{ \begin{array}{c} 0 \\ 0 \\ 0 \\ 0 \\ \mu^{\text{visc}}_k q_0 q_{0,i} + \mu^{\text{visc}}_{q_1} q_1 q_{1,i} \\ \mu^{\text{visc}}_k q_{0,i} \\ \mu^{\text{visc}}_\epsilon q_{1,i} \end{array} \right\} \tag{91}$$

$$\boldsymbol{\mathcal{S}} = \left\{ \begin{array}{c} 0 \\ \overline{\rho}\tilde{b}_1 \\ \overline{\rho}\tilde{b}_2 \\ \overline{\rho}\tilde{b}_3 \\ \tilde{u}_i\overline{\rho}\tilde{b}_i + \overline{\rho}\tilde{r} - \frac{\mu_T^{\text{visc}}}{Pr_b}\tilde{b}_{i,i} \\ -\frac{1}{q_0}\frac{\mu_T^{\text{visc}}}{Pr_b}\tilde{b}_{i,i} \\ 0 \end{array} \right\} + \left\{ \begin{array}{c} 0 \\ -[\frac{1}{3}\overline{\rho}(q_0^2 + q_1^2)]_{,1} \\ -[\frac{1}{3}\overline{\rho}(q_0^2 + q_1^2)]_{,2} \\ -[\frac{1}{3}\overline{\rho}(q_0^2 + q_1^2)]_{,3} \\ -[\frac{1}{3}\overline{\rho}(q_0^2 + q_1^2)\tilde{u}_i]_{,i} \\ \mathcal{S}_{q_0} \\ \mathcal{S}_{q_1} \end{array} \right\} \tag{92}$$

where

$$\begin{aligned} \tau_{ij} &= 2(\mu^{\text{visc}} + \mu_T^{\text{visc}})S_{ij}^d(\tilde{\boldsymbol{u}}) + \mu_b^{\text{visc}} S_{kk}(\tilde{\boldsymbol{u}})\delta_{ij} \\ S_{ij}(\boldsymbol{u}) &= \frac{1}{2}(u_{i,j} + u_{j,i}) \\ S_{ij}^d(\boldsymbol{u}) &= S_{ij}(\boldsymbol{u}) - \frac{1}{3}S_{kk}(\boldsymbol{u})\delta_{ij} \\ q_i^{\text{heat}} &= -(\kappa + \kappa_T)\widetilde{T}_{,i} \\ \mu_k^{\text{visc}} &= \mu^{\text{visc}} + \frac{\mu_T^{\text{visc}}}{\text{Pr}_k} \\ \mu_{q_1}^{\text{visc}} &= \mu^{\text{visc}} + \frac{\mu_T^{\text{visc}}}{\text{Pr}_{q_1}} \\ \mu_\epsilon^{\text{visc}} &= \mu^{\text{visc}} + \frac{\mu_T^{\text{visc}}}{\text{Pr}_\epsilon} \end{aligned} \tag{93}$$

and

$$\begin{aligned} \mathcal{S}_{q_0} &= \frac{2}{q_0}\mu_T^{\text{visc}} S_{ij}^d(\tilde{\boldsymbol{u}})\tilde{u}_{i,j} - \frac{1}{3}\overline{\rho}q_0\tilde{u}_{i,i} - \frac{1}{q_0}\overline{\rho}\epsilon + \mu_k^{\text{visc}}\frac{q_{0,i}q_{0,i}}{q_0} \\ \mathcal{S}_{q_1} &= \mathcal{P}_{q_1} - \mathcal{D}_{q_1} + \mathcal{R}_{q_1} \end{aligned} \tag{94}$$

where $\mathcal{P}_{q_1}$, $\mathcal{D}_{q_1}$ and $\mathcal{R}_{q_1}$ depend upon the turbulence model.

To obtain the set of entropy variables, consider the generalized entropy function

$$\mathcal{H}(\boldsymbol{U}) = \overline{\rho}\hat{s}(\boldsymbol{U}) \tag{95}$$

where the $\hat{s}$ is the entropy function defined from the averaged variables $\boldsymbol{U}$ of the system of equations via the Gibbs-like relation

$$\widetilde{T}d\hat{s} = d\tilde{e} + \overline{p}\; d\frac{1}{\overline{\rho}} \tag{96}$$

Remember here that the averaging operators, denoted by bars and tildes, stand for space-time averaging. (95) and (96) are form identical to the entropy function for the one-equation turbulence model, but there the entropy was defined from time-averaged variables. The entropy variables are defined by the change of variables

$$\boldsymbol{V}^T = \frac{\partial \mathcal{H}(\boldsymbol{U})}{\partial \boldsymbol{U}} \tag{97}$$

The following relations, derived from the differential (96),

$$\begin{aligned} \left(\frac{\partial \hat{s}}{\partial \tilde{e}}\right)_{1/\overline{\rho}} &= \frac{1}{\widetilde{\widetilde{T}}} \\ \left(\frac{\partial \hat{s}}{\partial \hat{v}}\right)_{\tilde{e}} &= \frac{\overline{p}}{\widetilde{\widetilde{T}}} \end{aligned} \tag{98}$$

are useful to reach the expression for the entropy variables

$$\boldsymbol{V} = \frac{1}{\widetilde{\widetilde{T}}} \left\{ \begin{matrix} \hat{\mu} - |\tilde{\boldsymbol{u}}|^2/2 - q_0^2/2 - q_1^2/2 \\ \tilde{u}_1 \\ \tilde{u}_2 \\ \tilde{u}_3 \\ -1 \\ q_0 \\ q_1 \end{matrix} \right\} \tag{99}$$

where $\hat{\mu} = \tilde{e} + \overline{p}/\overline{\rho} - \widetilde{\widetilde{T}}\hat{s}$ is the chemical potential per unit mass, defined from variables present in $\boldsymbol{U}$.

Under the change of variables $\boldsymbol{U} \mapsto \boldsymbol{V}$, the quasi-linear form of the system (88) can be expressed as

$$\boldsymbol{A}_0 \boldsymbol{V}_{,t} + \boldsymbol{A}_i \boldsymbol{V}_{,i} = (\boldsymbol{K}_{ij} \boldsymbol{V}_{,j})_{,i} + \boldsymbol{\mathcal{S}} \tag{100}$$

where

$$\boldsymbol{A}_0 = \boldsymbol{U}_{,\boldsymbol{V}} \tag{101}$$

$$\boldsymbol{A}_i = \boldsymbol{F}_{i,\boldsymbol{V}}^{\mathrm{adv}} \tag{102}$$

$$\boldsymbol{\mathcal{S}} = \boldsymbol{\mathcal{S}}(\boldsymbol{V}, \nabla \boldsymbol{V}) \tag{103}$$

and $\boldsymbol{K}_{ij}$ is formed such that

$$\boldsymbol{F}_i^{\mathrm{diff}} = \boldsymbol{K}_{ij} \boldsymbol{V}_j \tag{104}$$

Expressions for the coefficient matrices $\boldsymbol{A}_0$ and $\boldsymbol{A}_i$ can be found in Appendix A. Indeed, they possess these properties:

*i.* $\boldsymbol{A}_0$ is symmetric positive-definite.

*ii.* The $\boldsymbol{A}_i$'s are symmetric.

As a consequence, regarding the advective operator, $\mathcal{H}(\boldsymbol{U})$ is a *generalized entropy function.* However, we still need to examine under what conditions the symmetry of $\boldsymbol{K} = [\boldsymbol{K}_{ij}]$ is also attained.

From the diffusive flux vector (91) and the definition of $\boldsymbol{K}_{ij}$ (104), using the relations between the derivatives of primitive variables and entropy variables

$$\tilde{u}_{i,j} = \widetilde{T} V_{i+1,j} + \widetilde{T} \tilde{u}_i V_{5,j} \qquad \text{for } i = 1, 2, 3 \tag{105}$$

$$q_{i,j} = \widetilde{T} V_{i+6,j} + \widetilde{T} q_i V_{5,j} \qquad \text{for } i = 0, 1 \tag{106}$$

$$\widetilde{T}_{,i} = \widetilde{T}^2 V_{5,i} \tag{107}$$

we obtain, for instance,

$$\boldsymbol{K}_{11} = \widetilde{T} \begin{bmatrix} 0 & 0 & 0 & 0 & 0 & 0 & 0 \\ & \chi^{\text{visc}} & 0 & 0 & \chi^{\text{visc}} u_1 & 0 & 0 \\ & & \hat{\mu}^{\text{visc}} & 0 & \hat{\mu}^{\text{visc}} u_2 & 0 & 0 \\ & & & \hat{\mu}^{\text{visc}} & \hat{\mu}^{\text{visc}} u_3 & 0 & 0 \\ & & & & k_{55}^{11} & \mu_k^{\text{visc}} q_0 & \mu_{q_1}^{\text{visc}} q_1 \\ & \text{symm.} & & & \mu_k^{\text{visc}} q_0 & \mu_k^{\text{visc}} & 0 \\ & & & & \mu_\epsilon^{\text{visc}} q_1 & 0 & \mu_\epsilon^{\text{visc}} \end{bmatrix} \tag{108}$$

where definitions of $\hat{\mu}^{\text{visc}}$ and $k_{55}^{11}$ can be found in Appendix A along with the expressions for all the $\boldsymbol{K}_{ij}$s. Only $\boldsymbol{K}_{11}$, $\boldsymbol{K}_{22}$ and $\boldsymbol{K}_{33}$ need to be analyzed for symmetry. Then, to attain a symmetric $\boldsymbol{K} = [\boldsymbol{K}_{ij}]$ we have two possible assumptions:

*i.* We can choose

$$\mu_{q_1}^{\text{visc}} = \mu_\epsilon^{\text{visc}} \tag{109}$$

In this case, however, we are slightly deviating from the $k$-$\epsilon$ model assumption that $\mu_{q_1}^{\text{visc}} = \mu_k^{\text{visc}}$.

*ii.* If, according to the $k$-$\epsilon$ model and the space-time averaged system of equations (75), $\mu_{q_1}^{\text{visc}} = \mu_k^{\text{visc}}$, then symmetry requires that

$$\mu_\epsilon^{\text{visc}} = \mu_k^{\text{visc}} \tag{110}$$

Under this assumption, Mohammadi and Pironneau [33] showed that the $k$-$\epsilon$ model based on the variables $k$ and $\theta = k/\epsilon$ preserves positivity, i.e, $k$ and $\theta$ remain positive and therefore the system of equations is dissipative. This assumption is also satisfied by the $k$-$\omega$ model of Wilcox [34, 35] and the $k$-$\tau$ model of Speziale *et al.* [36].

Assumption *ii.*, however, would change the diffusive constant of the dissipation rate in the $k$-$\epsilon$ model, so the model constants would need to be retuned. Thus, assumption *i.* is going to be pursued while the other one is left for future research. In this case, $\boldsymbol{K}$ is also symmetric and positive-semidefinite. Appendix A includes the expression of the coefficient matrices that result from this development.

## 5. Clausius-Duhem inequality for the $k$-$\epsilon$ model

The Clausius-Duhem inequality is obtained by taking the dot product of the system of equations with the entropy variables

$$\boldsymbol{V} \cdot [\boldsymbol{A}_0 \boldsymbol{V}_{,t} + \boldsymbol{A}_i \boldsymbol{V}_{,i} = (\boldsymbol{K}_{ij} \boldsymbol{V}_{,j})_{,i} + \boldsymbol{\mathcal{S}}] \tag{111}$$

Applying the chain rule to the diffusive term and substituting the definition of the fluxes

$$\boldsymbol{V} \cdot \boldsymbol{A}_0 \boldsymbol{V}_{,t} + \boldsymbol{V} \cdot \boldsymbol{F}^{\text{adv}}_{i,i} - \left(\boldsymbol{V} \cdot \boldsymbol{F}^{\text{diff}}_i\right)_{,i} - \boldsymbol{V} \cdot \boldsymbol{\mathcal{S}} = -\boldsymbol{V}_{,i} \cdot \boldsymbol{K}_{ij} \boldsymbol{V}_{,j} \tag{112}$$

Let us examine with detail the different terms present in (112).

$$\boldsymbol{V} \cdot \boldsymbol{A}_0 \boldsymbol{V}_{,t} = \mathcal{H}_{,t} \tag{113}$$

$$\boldsymbol{V} \cdot \boldsymbol{F}^{\text{adv}}_{i,i} = [\mathcal{H} \tilde{u}_i]_{,i} \tag{114}$$

$$\boldsymbol{V} \cdot \boldsymbol{F}^{\text{diff}}_i = -\frac{(\kappa + \kappa_T)\widetilde{T}_{,i}}{\widetilde{T}} + \left(\mu^{\text{visc}}_{\epsilon} - \mu^{\text{visc}}_{q_1}\right) \frac{q_1 q_{1,i}}{\widetilde{T}} \tag{115}$$

$$
\begin{aligned}
\boldsymbol{V}_{,i} \cdot \boldsymbol{K}_{ij} \boldsymbol{V}_{,j} &= \boldsymbol{V}_{,i} \cdot \boldsymbol{F}_i^{\text{diff}} \\
&= \left( \frac{\tilde{u}_{j,i}}{\widetilde{T}} - \frac{\tilde{u}_j \widetilde{T}_{,i}}{\widetilde{T}^2} \right) \left[ 2(\mu^{\text{visc}} + \mu_T^{\text{visc}}) S_{ij}^d(\tilde{\boldsymbol{u}}) + \mu_b^{\text{visc}} S_{kk} \delta_{ij} \right] \\
&+ \left( \frac{\widetilde{T}_{,i}}{\widetilde{T}^2} \right) \left[ 2(\mu^{\text{visc}} + \mu_T^{\text{visc}}) S_{ij}^d(\tilde{\boldsymbol{u}}) + \mu_b^{\text{visc}} S_{kk} \delta_{ij} \right] \tilde{u}_j \\
&+ \left( \frac{\widetilde{T}_{,i}}{\widetilde{T}^2} \right) \left[ (\kappa + \kappa_T) \widetilde{T}_{,i} + \mu_k^{\text{visc}} q_0 q_{0,i} + \mu_{q_1}^{\text{visc}} q_1 q_{1,i} \right] \\
&+ \left( \frac{q_{0,i}}{\widetilde{T}} - \frac{q_0 \widetilde{T}_{,i}}{\widetilde{T}^2} \right) \left[ \mu_k^{\text{visc}} q_{0,i} \right] \\
&+ \left( \frac{q_{1,i}}{\widetilde{T}} - \frac{q_1 \widetilde{T}_{,i}}{\widetilde{T}^2} \right) \left[ \mu_\epsilon^{\text{visc}} q_{1,i} \right]
\end{aligned} \tag{116}
$$

$$
\begin{aligned}
&= \left( \frac{\tilde{u}_{j,i}}{\widetilde{T}} \right) \left[ 2(\mu^{\text{visc}} + \mu_T^{\text{visc}}) S_{ij}^d(\tilde{\boldsymbol{u}}) + \mu_b^{\text{visc}} S_{kk} \delta_{ij} \right] \\
&+ \left( \frac{\widetilde{T}_{,i}}{\widetilde{T}^2} \right) (\kappa + \kappa_T) \widetilde{T}_{,i} \\
&+ \left( \frac{q_{0,i}}{\widetilde{T}} \right) \mu_k^{\text{visc}} q_{0,i} \\
&+ \left( \frac{q_{1,i}}{\widetilde{T}} \right) \mu_\epsilon^{\text{visc}} q_{1,i} - \left( \frac{\widetilde{T}_{,i}}{\widetilde{T}^2} \right) \left[ \mu_\epsilon^{\text{visc}} - \mu_{q_1}^{\text{visc}} \right] q_1 q_{1,i}
\end{aligned} \tag{117}
$$

$$
\begin{aligned}
&= \frac{\Upsilon(\tilde{\boldsymbol{u}}, \tilde{\boldsymbol{u}})}{\widetilde{T}} + (\kappa + \kappa_T) \frac{\widetilde{T}_{,i} \widetilde{T}_{,i}}{\widetilde{T}^2} \\
&+ 2\mu_T^{\text{visc}} \frac{S_{ij}^d(\tilde{\boldsymbol{u}}) \tilde{u}_{i,j}}{\widetilde{T}} + \mu_k^{\text{visc}} \frac{q_{0,i} q_{0,i}}{\widetilde{T}} + \mu_\epsilon^{\text{visc}} \frac{q_{1,i} q_{1,i}}{\widetilde{T}} \\
&+ \left( \frac{\widetilde{T}_{,i}}{\widetilde{T}^2} \right) \left[ \mu_{q_1}^{\text{visc}} - \mu_\epsilon^{\text{visc}} \right] q_1 q_{1,i}
\end{aligned} \tag{118}
$$

$$
\begin{aligned}
\boldsymbol{V} \cdot \boldsymbol{\mathcal{S}} &= \left( \frac{\tilde{u}_i}{\widetilde{T}} \right) \left[ -\frac{1}{3} \overline{\rho} (q_0^2 + q_1^2) \right]_{,i} \\
&- \left( \frac{1}{\widetilde{T}} \right) \left[ \overline{\rho} \tilde{r} - \left( \frac{1}{3} \overline{\rho} (q_0^2 + q_1^2) \tilde{u}_i \right)_{,i} \right] \\
&+ \left( \frac{q_0}{\widetilde{T}} \right) \left[ \frac{2}{q_0} \mu_T^{\text{visc}} S_{ij}^d(\tilde{\boldsymbol{u}}) \tilde{u}_{i,j} - \frac{1}{3} \overline{\rho} q_0 \tilde{u}_{i,i} - \frac{1}{q_0} \overline{\rho} \epsilon + \mu_k^{\text{visc}} \frac{q_{0,i} q_{0,i}}{q_0} \right] \\
&+ \left( \frac{q_1}{\widetilde{T}} \right) \mathcal{S}_{q_1}
\end{aligned} \tag{119}
$$

Gathering all the terms results in the Clausius-Duhem equality for the modeled

Reynolds-averaged equations coupled with a two-equation model,

$$
\begin{aligned}
\mathcal{H}_{,t} + [\mathcal{H}\tilde{u}_i]_{,i} - \left( \frac{-(\kappa+\kappa_T)\widetilde{T}_{,i}}{\widetilde{T}} \right)_{,i} + \frac{\overline{\rho}\tilde{r}}{\widetilde{T}} = & - \left( \frac{\Upsilon(\tilde{\boldsymbol{u}}, \tilde{\boldsymbol{u}})}{\widetilde{T}} \right) - (\kappa+\kappa_T) \frac{\widetilde{T}_{,i}\widetilde{T}_{,i}}{T^2} - \frac{\overline{\rho}\epsilon}{\widetilde{T}} \\
& + \left( (\mu_\epsilon^{\text{visc}} - \mu_{q_1}^{\text{visc}}) \frac{q_1 q_{1,i}}{\widetilde{T}} \right)_{,i} \\
& - \frac{\widetilde{T}_{,i}}{\widetilde{T}} (\mu_\epsilon^{\text{visc}} - \mu_{q_1}^{\text{visc}}) \frac{q_1 q_{1,i}}{\widetilde{T}} \\
& + \frac{1}{\widetilde{T}} \frac{1}{3} \overline{\rho} q_1^2 \tilde{u}_{i,i} - \mu_\epsilon^{\text{visc}} \frac{q_{1,i} q_{1,i}}{\widetilde{T}} \\
& + \left( \frac{q_1}{\widetilde{T}} \right) \mathcal{S}_{q_1}
\end{aligned}
\tag{120}
$$

### 5.1. Consequences for modeling

Equation (120) is the entropy production equality for the modeled system of equations. For positive generalized entropy production it is necessary that the right-hand-side of (120) is negative. The terms multiplying the factor $(\mu_\epsilon^{\text{visc}} - \mu_{q_1}^{\text{visc}})$ are indefinite because they can have either sign. For entropy production then, we can set

$$
\mu_\epsilon^{\text{visc}} = \mu_{q_1}^{\text{visc}} \tag{121}
$$

which in turn implies $Pr_\epsilon = Pr_{q_1}$. If according to the assumptions for the $k$-$\epsilon$ model $\mu_k^{\text{visc}} = \mu_{q_1}^{\text{visc}}$, then

$$
\mu_\epsilon^{\text{visc}} = \mu_k^{\text{visc}} \tag{122}
$$

or $Pr_\epsilon = Pr_k$. This result was already pointed out by Mohammadi and Pironneau [33] for their positivity and stability analysis of the $k$-$\epsilon$ model.

For entropy production we may further impose to the model

$$
\frac{\overline{\rho}\epsilon}{q_1} - \mathcal{S}_{q_1} - \frac{1}{3}\overline{\rho} q_1 \tilde{u}_{i,i} + \mu_\epsilon^{\text{visc}} \frac{q_{1,i} q_{1,i}}{q_1} \geq 0 \tag{123}
$$

or the more restricting condition,

$$
-\mathcal{S}_{q_1} - \frac{1}{3}\overline{\rho} q_1 \tilde{u}_{i,i} + \mu_\epsilon^{\text{visc}} \frac{q_{1,i} q_{1,i}}{q_1} \geq 0 \tag{124}
$$

which can also be extended to any two-equation turbulence model. These assumptions should be further examined with DNS data. Under these hypotheses, the Clausius-Duhem equality becomes

$$
\boxed{
\begin{aligned}
(\overline{\rho}\hat{s})_{,t} + (\overline{\rho}\hat{s}\tilde{u}_i)_{,i} + \left( \frac{-(\kappa+\kappa_T)\widetilde{T}_{,i}}{\widetilde{T}} \right)_{,i} &= \frac{\Upsilon(\tilde{\boldsymbol{u}},\tilde{\boldsymbol{u}})}{\widetilde{T}} + (\kappa+\kappa_T)\,\frac{\widetilde{T}_{,i}\widetilde{T}_{,i}}{\widetilde{T}^2} + \frac{\overline{\rho}\epsilon}{\widetilde{T}} \\
&\quad - \frac{1}{\widetilde{T}}\frac{1}{3}\overline{\rho}q_1^2\tilde{u}_{i,i} + \mu_\epsilon^{\text{visc}}\frac{q_{1,i}q_{1,i}}{\widetilde{T}} \\
&\quad - \left(\frac{q_1}{\widetilde{T}}\right)\mathcal{S}_{q_1} \\
&\geq 0
\end{aligned}
}
\tag{125}
$$

Thus, this can be satisfied by a modeled $q_1$ equation with a source term of the form

$$
\mathcal{S}_{q_1} = \mathcal{P}_{q_1} - \frac{1}{3}\overline{\rho}q_1\tilde{u}_{i,i} + \mu_\epsilon^{\text{visc}}\frac{q_{1,i}q_{1,i}}{q_1} - \mathcal{D}_{q_1} \tag{126}
$$

where $\mathcal{P}_{q_1}$ and $\mathcal{D}_{q_1}$ are the production and the dissipation of $q_1$, respectively, such that

$$
\mathcal{D}_{q_1} \geq \mathcal{P}_{q_1} \tag{127}
$$

## 6. Application to the $k$-$\epsilon$ model

In previous sections, we have introduced the new variables $q_0$ and $q_1$ that symmetrize the coupled system for the $k$-$\epsilon$ model. The equation for $q_0$ has already been presented in Section 3.2. Now we turn our attention to the dissipation rate equation, $\epsilon^*$, which is transformed via the change of variables

$$
q_1^4 = \nu\epsilon^* \tag{128}
$$

into

$$
[\overline{\rho}q_1]_{,t} + [\overline{\rho}\tilde{u}_i q_1]_{,i} = [(\mu^{\text{visc}} + \frac{\mu_T^{\text{visc}}}{\text{Pr}_\epsilon})q_{1,i}]_{,i} + \mathcal{S}_{q_1} \tag{129}
$$

Neglecting variations of viscosity,

$$
4q_1^3 dq_1 = \nu d\epsilon^* \tag{130}
$$

$$
dq_1 = \frac{q_1}{4\epsilon^*}d\epsilon^* \tag{131}
$$

Hence, the source term for the $q_1$ equation is as a function of $\mathcal{S}_\epsilon$

$$\mathcal{S}_{q_1} = \mathcal{P}_{q_1} - \mathcal{D}_{q_1} + \mathcal{R}_{q_1} \tag{132}$$

$$= \frac{q_1}{4\epsilon^*}\mathcal{S}_\epsilon + 3\mu_\epsilon^{\text{visc}} \frac{q_{1,i}q_{1,i}}{q_1} \tag{133}$$

where

$$\mathcal{S}_\epsilon = \mathcal{P}_\epsilon - \mathcal{D}_\epsilon + \mathcal{R}_\epsilon \tag{134}$$

The source terms for $q_1$ are defined in relation to the source terms of $\epsilon$ as

$$\begin{aligned} \mathcal{P}_{q_1} &= \frac{q_1}{4\epsilon^*}\mathcal{P}_\epsilon \\ \mathcal{D}_{q_1} &= \frac{q_1}{4\epsilon^*}\mathcal{D}_\epsilon \\ \mathcal{R}_{q_1} &= \frac{q_1}{4\epsilon^*}\mathcal{R}_\epsilon + 3\mu_\epsilon^{\text{visc}} \frac{q_{1,i}q_{1,i}}{q_1} \end{aligned} \tag{135}$$

Thus,

$$\begin{aligned} \mathcal{P}_{q_1} &= \frac{C_{\epsilon 1} f_1}{4} \frac{q_1}{k} \left( \mathcal{P}_k + \frac{2}{3}\overline{\rho}\tilde{u}_{i,i} k \right) \\ \mathcal{D}_{q_1} &= \frac{C_{\epsilon 2} f_2}{4} \overline{\rho} \frac{q_1^5}{\nu k} \\ \mathcal{R}_{q_1} &= \frac{q_1}{4\epsilon^*}\mathcal{R}_\epsilon + 3\mu_\epsilon^{\text{visc}} \frac{q_{1,i}q_{1,i}}{q_1} \end{aligned} \tag{136}$$

and the eddy viscosity transforms under the change of variables into

$$\mu_T^{\text{visc}} = \frac{C_\mu}{4} f_\mu \overline{\rho} \left( \frac{q_0}{q_1} \right)^4 \tag{137}$$

From the strict entropy production condition (124), the source term needs to satisfy

$$-\mathcal{S}_{q_1} + \mu_\epsilon^{\text{visc}} \frac{q_{1,i}q_{1,i}}{q_1} - \frac{1}{3}\overline{\rho} q_1 \tilde{u}_{i,i} \geq 0 \tag{138}$$

which is imposed by increasing $\mathcal{D}_{q_1}$ until the equality is met:

$$\mathcal{D}_{q_1} = \max(\mathcal{D}_{q_1}, \mathcal{P}_{q_1} + \mathcal{R}_{q_1} - \mu_\epsilon^{\text{visc}} \frac{q_{1,i}q_{1,i}}{q_1} + \frac{1}{3}\overline{\rho} q_1 \tilde{u}_{i,i}) \tag{139}$$

For the case of the Lam-Bremhorst $k$-$\epsilon$ model [37], the extra source term is zero,

$$\mathcal{R}_\epsilon = 0$$

and the damping functions are given by

$$
\begin{aligned}
f_\mu &= \left(1.0 - e^{-0.0165 Re_y}\right)^2 \left(1.0 + \frac{20.5}{Re_T}\right) \\
f_1 &= 1.0 + \left(\frac{0.05}{f_\mu}\right)^3 \\
f_2 &= 1.0 - e^{-Re_T^2}
\end{aligned}
\tag{140}
$$

where the turbulent Reynolds numbers are

$$
Re_T = \frac{1}{4}\left(\frac{q_0}{q_1}\right)^4 \qquad Re_y = \frac{\bar{\rho} q_0 y}{\sqrt{2}\ \mu^{\text{visc}}} \tag{141}
$$

and $y$ is the distance normal to the wall. The model constants are given in Table 6.1.

*Table 6.1.* Model constants.

| Model | $C_\mu$ | $C_{\epsilon 1}$ | $C_{\epsilon 2}$ | $Pr_k$ | $Pr_\epsilon$ |
|---|---|---|---|---|---|
| Lam-Bremhorst | 0.09 | 1.44 | 1.92 | 1.0 | 1.3 |

Thus, the strict entropy condition (138) implies for the Lam-Bremhorst model that

$$
-\mathcal{P}_{q_1} + \mathcal{D}_{q_1} - 2\mu_\epsilon^{\text{visc}} \frac{q_{1,i} q_{1,i}}{q_1} - \frac{1}{3}\bar{\rho} q_1 \tilde{u}_{i,i} \geq 0 \tag{142}
$$

Recall that we are employing the assumption for the turbulent transport of $k$, equation (67), with

$$
Pr_{q_1} = Pr_\epsilon \tag{143}
$$

## 7. Numerical implementation

The Reynolds-averaged Navier-Stokes equations have been implemented coupled with the $q_0$, $q_1$ equations and solved via a time-discontinuous Galerkin/least-squares method. The finite element algorithm follows along the lines described in Shakib *et al.* [38]. The solution is advanced in time via a predictor multicorrector algorithm and the linear system of equations is solved at each iteration with the GMRES solver [39].

The $k$-$\epsilon$ modeled equations are known to be a very stiff and unstable dynamical system. Usually the equation for $\epsilon$ has been blamed for the numerical difficulties encountered but in our implementation the $q_1$ equation exhibited very good stability in the numerical computations. In the $q_0$ equation, the source terms become unbounded and non-integrable as $q_0 \to 0$, unless the cross-diffusion term balances the dissipation term,

$$\frac{\overline{\rho} q_1^4}{\nu q_0} = \mu_k^{\mathrm{visc}} \frac{q_{0,k} q_{0,k}}{q_0} \tag{144}$$

This can be obviated by imposing (144) as the boundary condition near the wall, which was implemented via a penalty term for $y^+ < 3$.

## 8. Turbulence simulations

Several illustrative computations performed with the Lam-Bremhorst $k$-$\epsilon$ [37] model developed in Section 6 are presented. The main goal of this section is to validate the model and the assumptions introduced. It will be shown that solutions obtained with this model are very close to those obtained by standard versions of $k$-$\epsilon$ models, with the added nonlinear stability advantage that the second law of thermodynamics is embedded into the discrete formulation.

Since only steady state solutions will be computed, the solution is advanced in time with the constant-in-time algorithm, see [38]. The elements are bilinear quadrilaterals in space and are integrated with the usual $2 \times 2$ Gaussian quadrature rule.

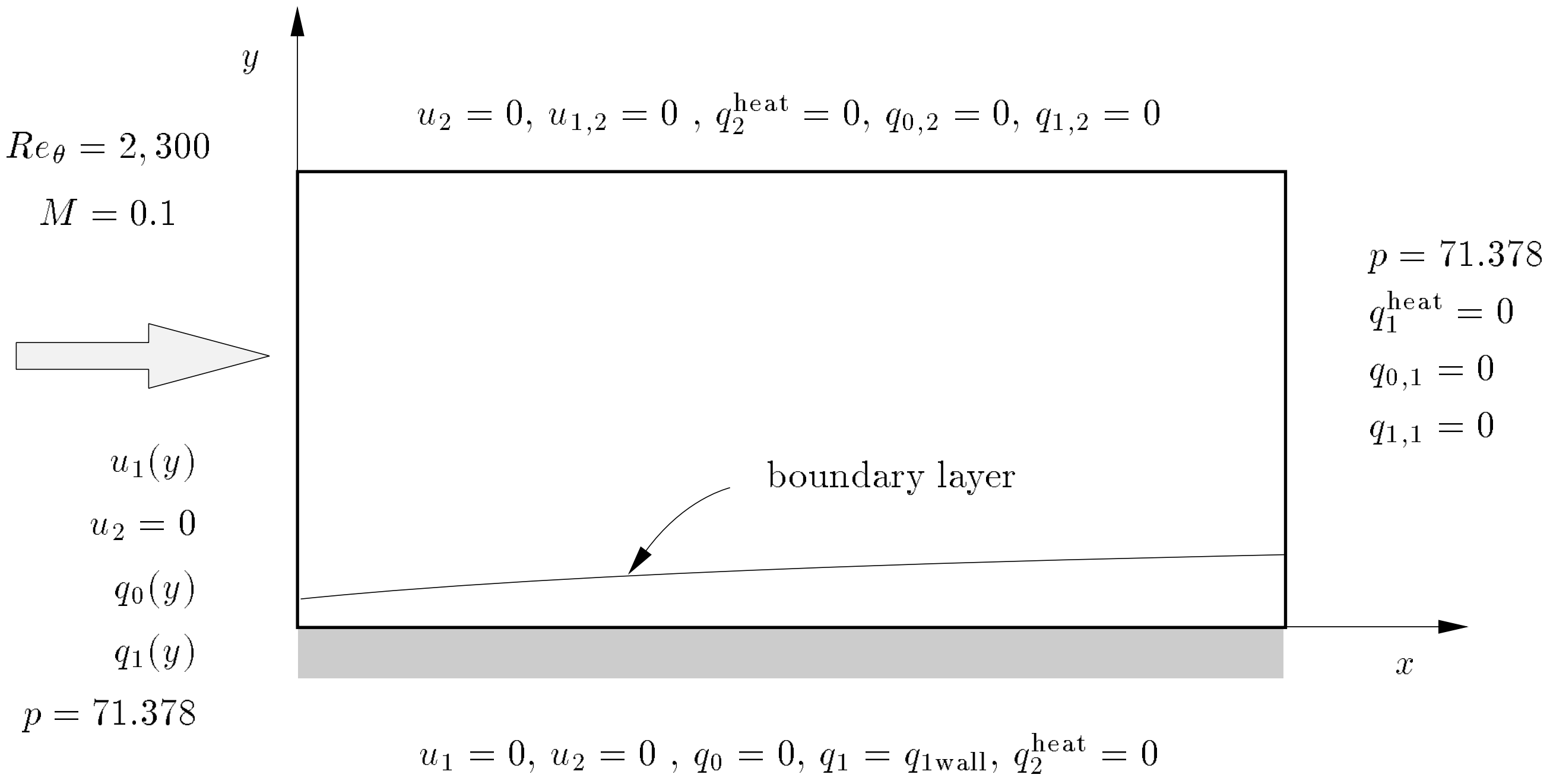

*Figure 2.* Turbulent flat plate problem.

### 8.1. Nearly incompressible flat plate boundary layer

Although the flow over a flat plate with constant pressure is one of the simplest flows, any promising turbulence model has to replicate it. To avoid the difficulties of transition from laminar to turbulent flow the plate is started at a Reynolds number based on momentum thickness of $Re_\theta = 2,300$. The momentum thickness, $\theta$, is defined as the integral

$$\theta = \int_0^{y_e} \frac{\overline{\rho}\tilde{u}_1(\tilde{u}_1^e - \tilde{u}_1)}{\overline{\rho}^e(\tilde{u}_1^e)^2} dy \tag{145}$$

where the superscript $e$ denotes the edge of the boundary layer. $\theta$ defines the corresponding Reynolds number

$$Re_\theta = \frac{\overline{\rho}\tilde{u}_1^e\theta}{\mu^{\rm visc}} \tag{146}$$

See White [40] for background. The free stream Mach number was 0.1 and inlet boundary conditions were obtained from another computation. Boundary conditions are depicted in Figure 2. In particular, $q_1$ is given by the balance of dissipation and cross-diffusion near the wall,

$$q_{1\rm wall}^4 = \nu^2 q_{0,k} q_{0,k} \tag{147}$$

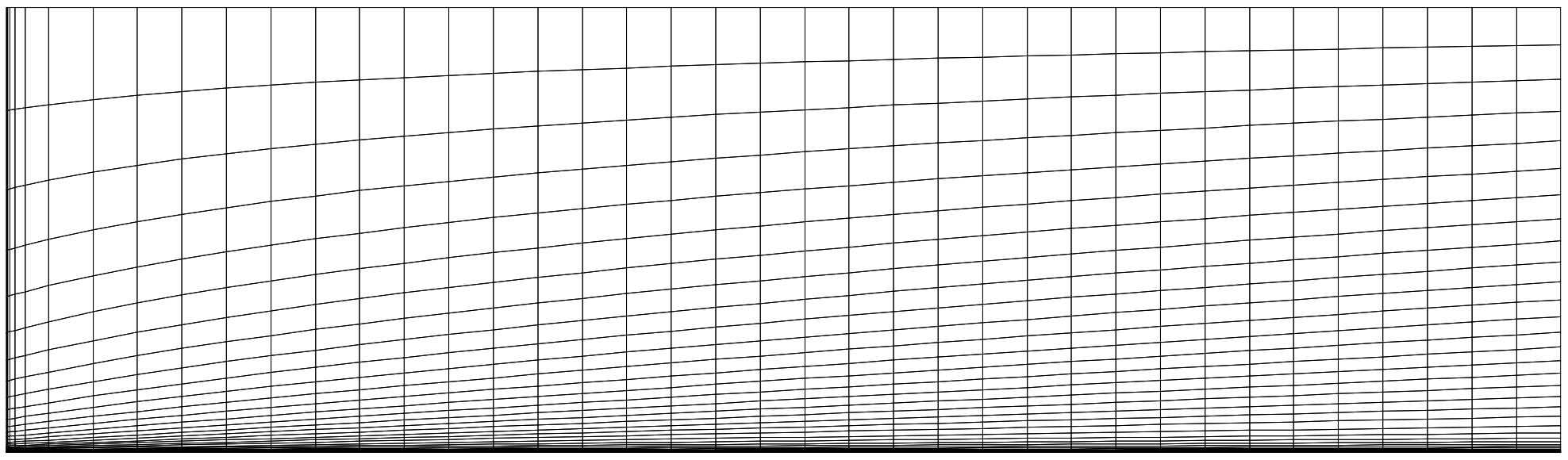

*Figure 3.* Turbulent flat plate. Mesh of $40 \times 60$ elements.

The domain, of dimensions $3.5 \times 1$, was discretized with $40 \times 60$ elements. The mesh, shown in Figure 3, accommodates the boundary layer growth, keeping a constant number of elements inside the boundary layer (40 elements) and a constant near-wall element size ($y^+ = 0.01$). The superscript + indicates wall coordinates, which are obtained by non-dimensionalizing the variables with the viscosity $\mu^{\mathrm{visc}}$ and the friction velocity $u_\tau$,

$$u_\tau = \sqrt{\frac{\tau_w}{\overline{\rho}}} \tag{148}$$

where $\tau_w$ is the shear stress at the wall. Thus,

$$\begin{aligned} y^+ &= \frac{\overline{\rho} u_\tau y}{\mu^{\mathrm{visc}}} = \frac{u_\tau y}{\nu} \\ u^+ &= \frac{u}{u_\tau} \end{aligned} \tag{149}$$

The friction coefficient is shown in Figure 4. It is compared to the computations realized with the Lam-Bremhorst $k$-$\epsilon$ model by Patel *et al.* [41] and Wilcox [34, 42]. The empirical correlation is given by that of Karman-Schönherr,

$$\frac{1}{C_f} = 17.08(\log_{10} Re_\theta)^2 + 25.11 \log_{10} Re_\theta + 6.012 \tag{150}$$

obtained from a wide set of experimental data (see Hopkins *et al.* [43]). After the quick transient near the inlet plane, the agreement of the present results and other computations is good. As observed for $k$-$\epsilon$ models, the friction coefficient is overestimated for the flat plate.

In Figure 5, velocity profiles at several values of $Re_\theta$ are plotted in log-log coordinates along with analytical asymptotic behavior, i.e., the law of the wall,

$$u^+ = y^+ \tag{151}$$

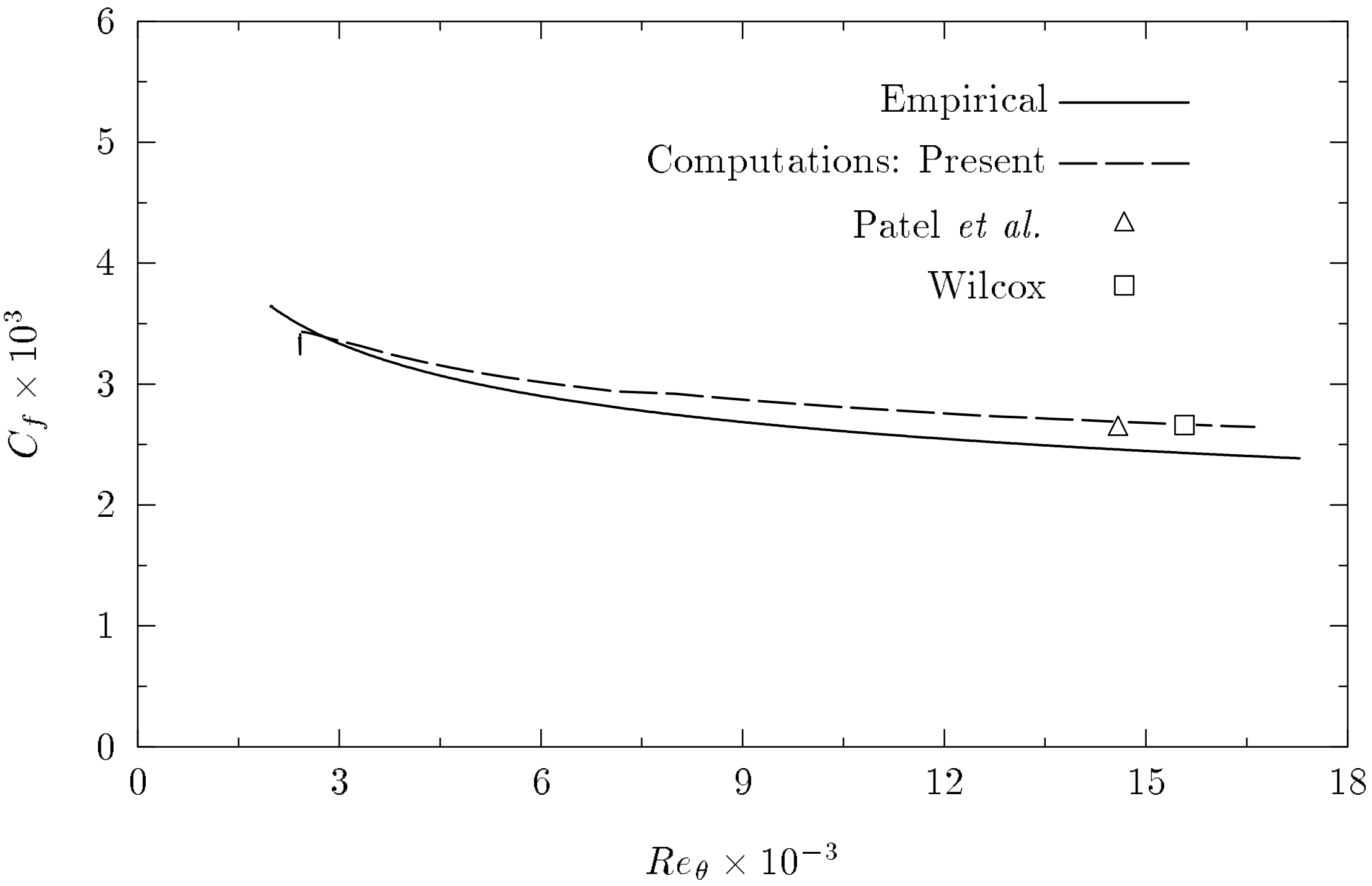

*Figure 4.* Flow over a flat plate. Friction coefficient for the Lam-Bremhorst model compared with empirical correlation and other researcher's computations. For data, see Patel et al. [41] and Wilcox [34, 42].

and the logarithmic law,

$$u^+ = \frac{1}{\mathcal{K}} \ln y^+ + B \tag{152}$$

where the following values have been employed: $\mathcal{K} = 0.41$ and $B = 5.2$. The results are similar to those obtained with the Lam-Bremhorst model (see Patel *et al.* [41] and Wilcox [42]). Note that the velocity profiles follow very closely the law of the wall, but are very slightly underestimated in the logarithmic region.

Figure 6 shows the turbulent velocity scale in wall coordinates,

$$q_0^+ = \frac{q_0}{u_\tau} \tag{153}$$

Indeed, $q_0^+$ is linear near the wall and presents a maximum at about $y^+ = 20.$ where $q_0^+ = 3.$, corresponding to $k^+ = 4.5$. These values are in very good accordance with the experimental data reported by Patel *et al.* [41], where representative values are $k^+ = 4.5$ at $y^+ = 15$. Note, though, that the computed peak of $q_0^+$ occurs at a slightly larger $y^+$. The turbulent velocity scale decays to the free stream value near the edge of the boundary layer.

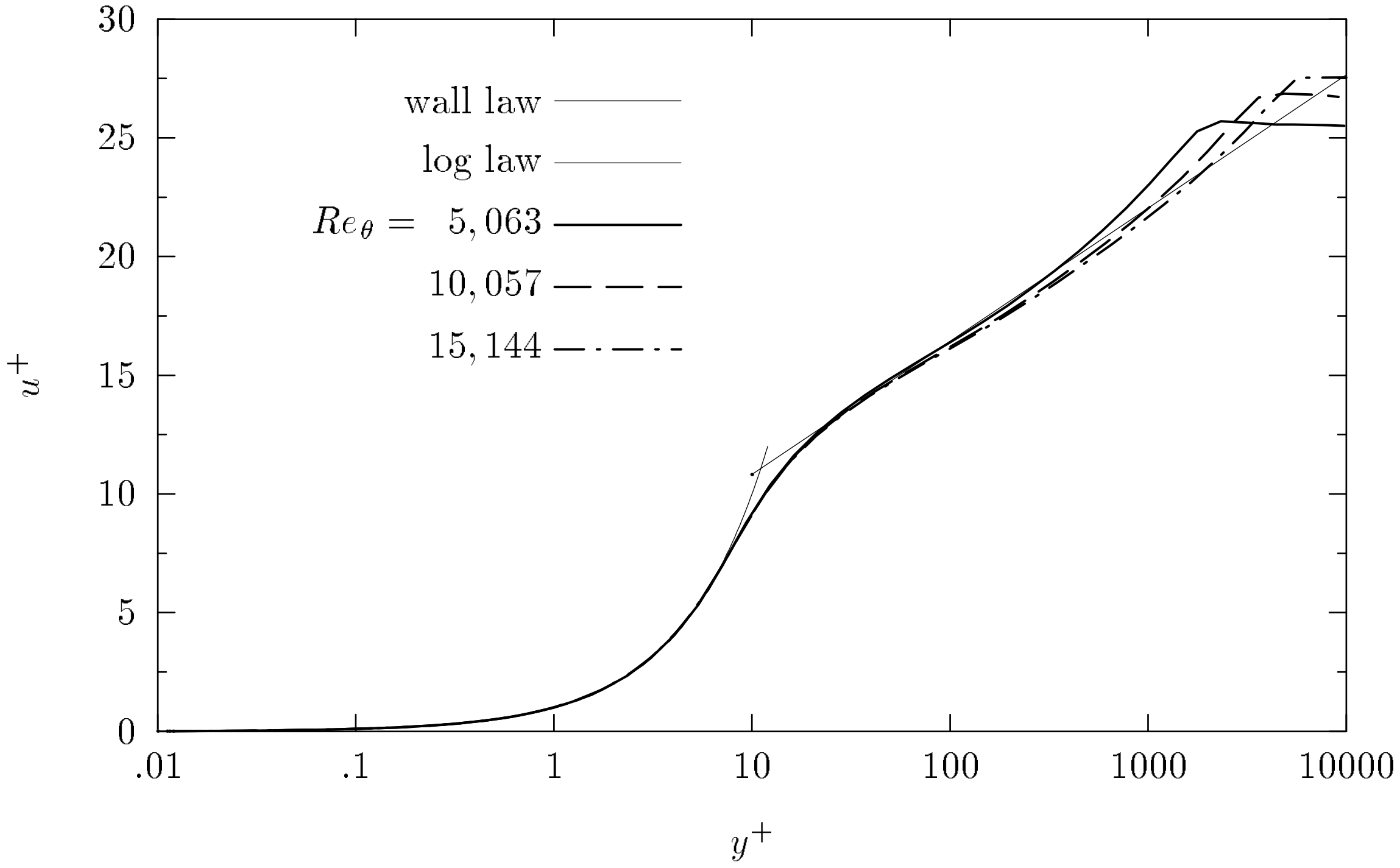

*Figure 5.* Flow over a flat plate. Velocity profiles in wall coordinates for several values of $Re_\theta$.

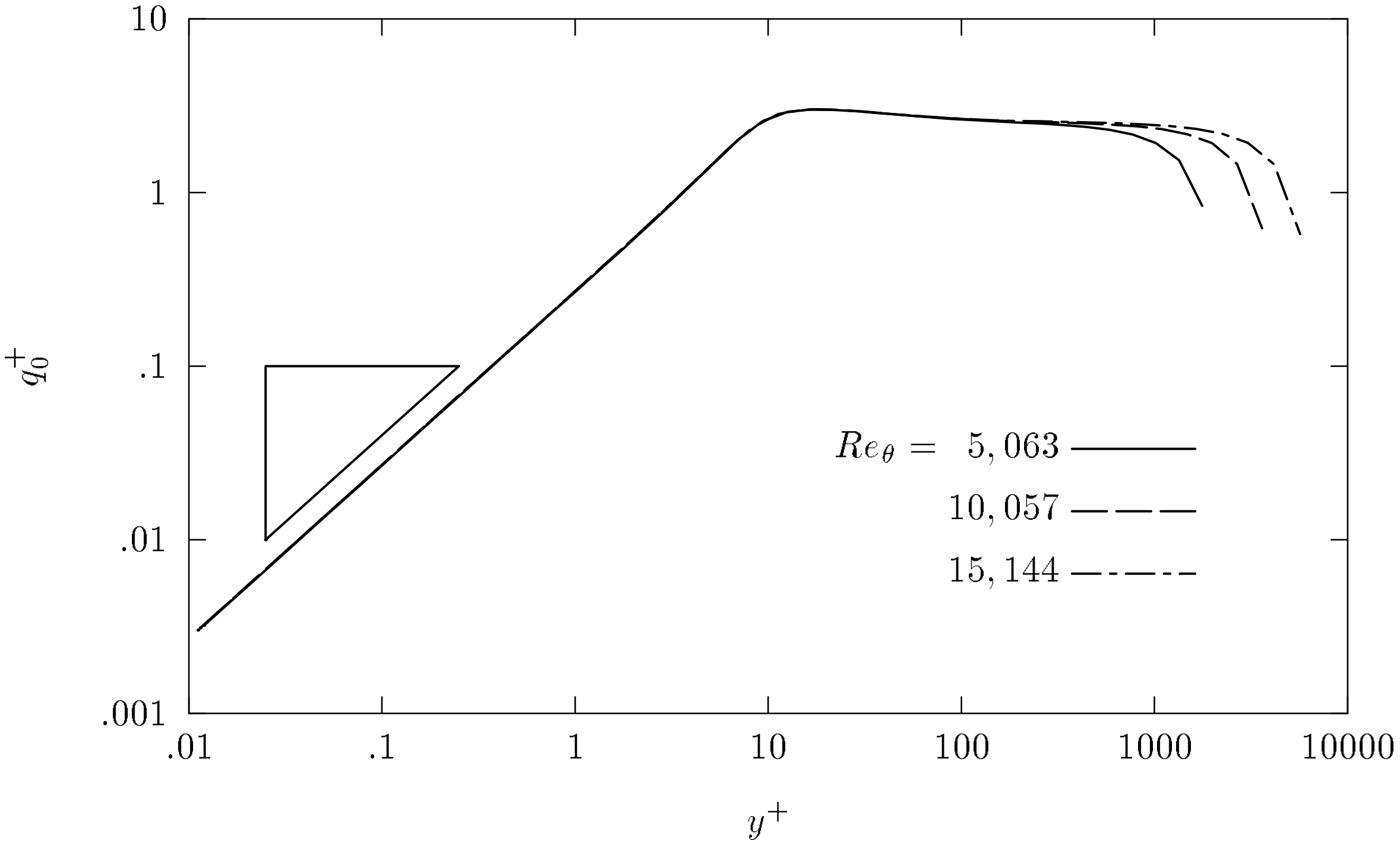

*Figure 6.* Flow over a flat plate. Turbulent velocity scale in wall coordinates for several values of $Re_\theta$.

The source budgets of $q_0$ and $k$ in Figures 7 and 8, respectively, show the usual characteristics (see Mansour *et al.* [44]). Despite being singular, dissipation and cross-diffusion of $q_0$ balance near the wall. This is achieved through the boundary condition for $q_1$. In the log region, $y^+ > 50$, production and dissipation balance each other. The diffusion of $q_0$ vanishes at the wall. The budget for $k$ shows that near-wall diffusion and dissipation balance and likewise production and dissipation in the logarithmic region.

The turbulent shear stress is shown in Figure 9, and Figure 10 displays the modeling functions of the Lam-Bremhorst turbulence model inside the boundary layer.

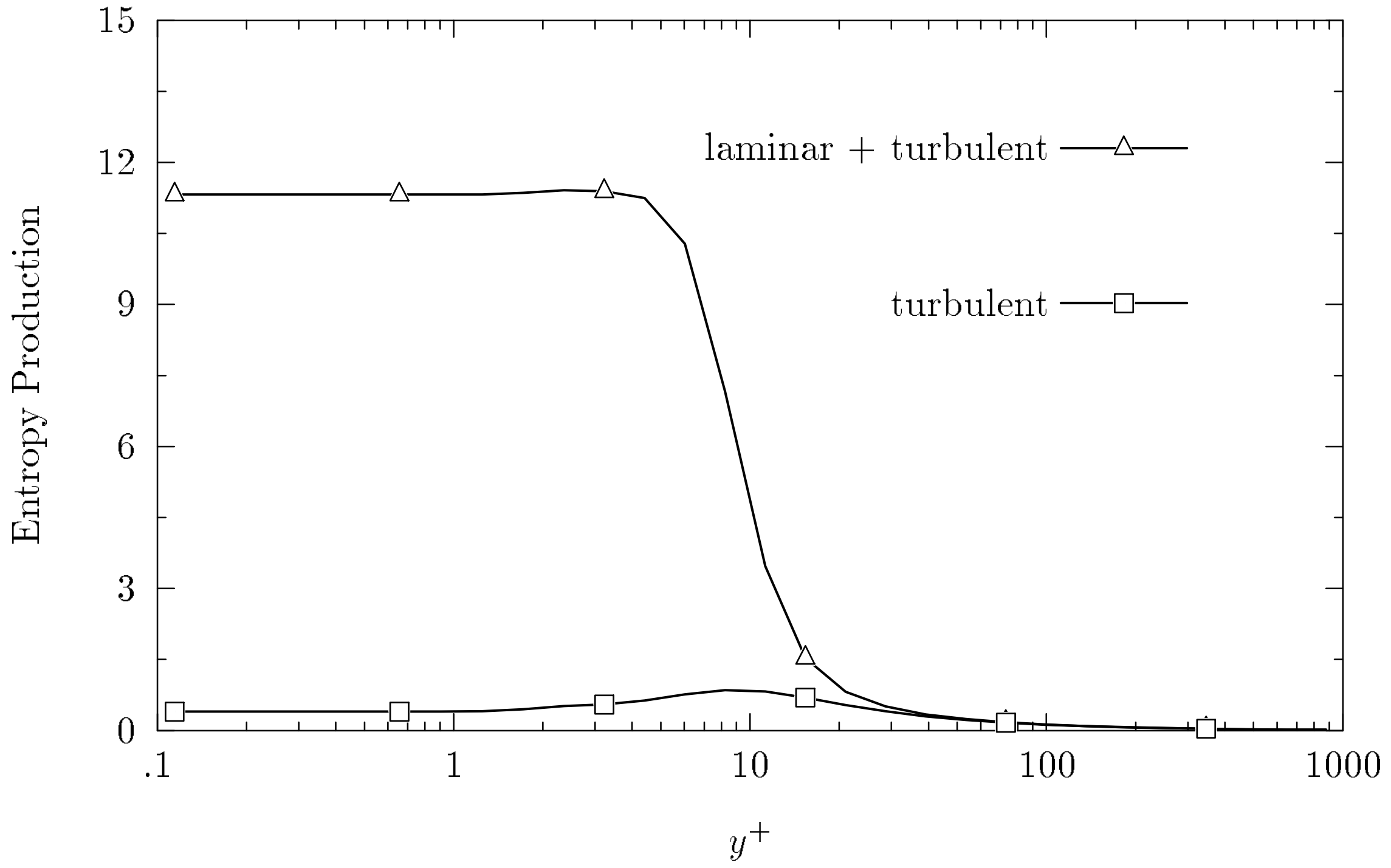

*Figure 11.* Flow over a flat plate. Isothermal entropy production at $Re_\theta = 15,144$.

Figure 11 depicts the entropy production across the boundary layer, showing the turbulence contribution for the Lam-Bremhorst model,

$$\overline{\rho}\epsilon + q_1(\mathcal{D}_{q_1} - \mathcal{P}_{q_1} - 2/3\mathcal{R}_{q_1}) \tag{154}$$

and the total entropy production,

$$\Upsilon(\tilde{\boldsymbol{u}}, \tilde{\boldsymbol{u}}) + \overline{\rho}\epsilon + q_1(\mathcal{D}_{q_1} - \mathcal{P}_{q_1} - 2/3\mathcal{R}_{q_1}) \tag{155}$$

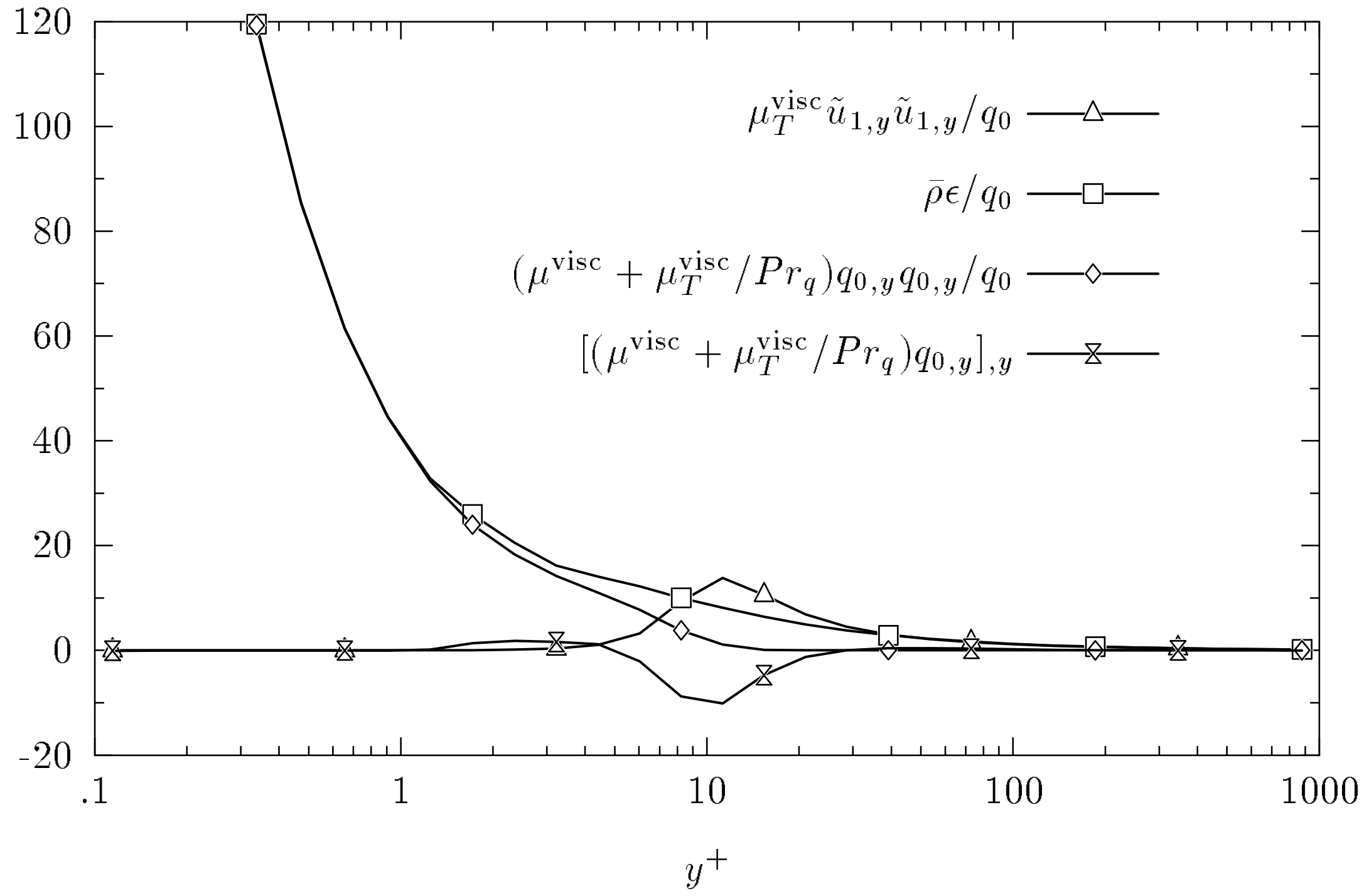

*Figure 7.* Flow over a flat plate. Budget for $q_0$ at $Re_\theta = 15,144$.

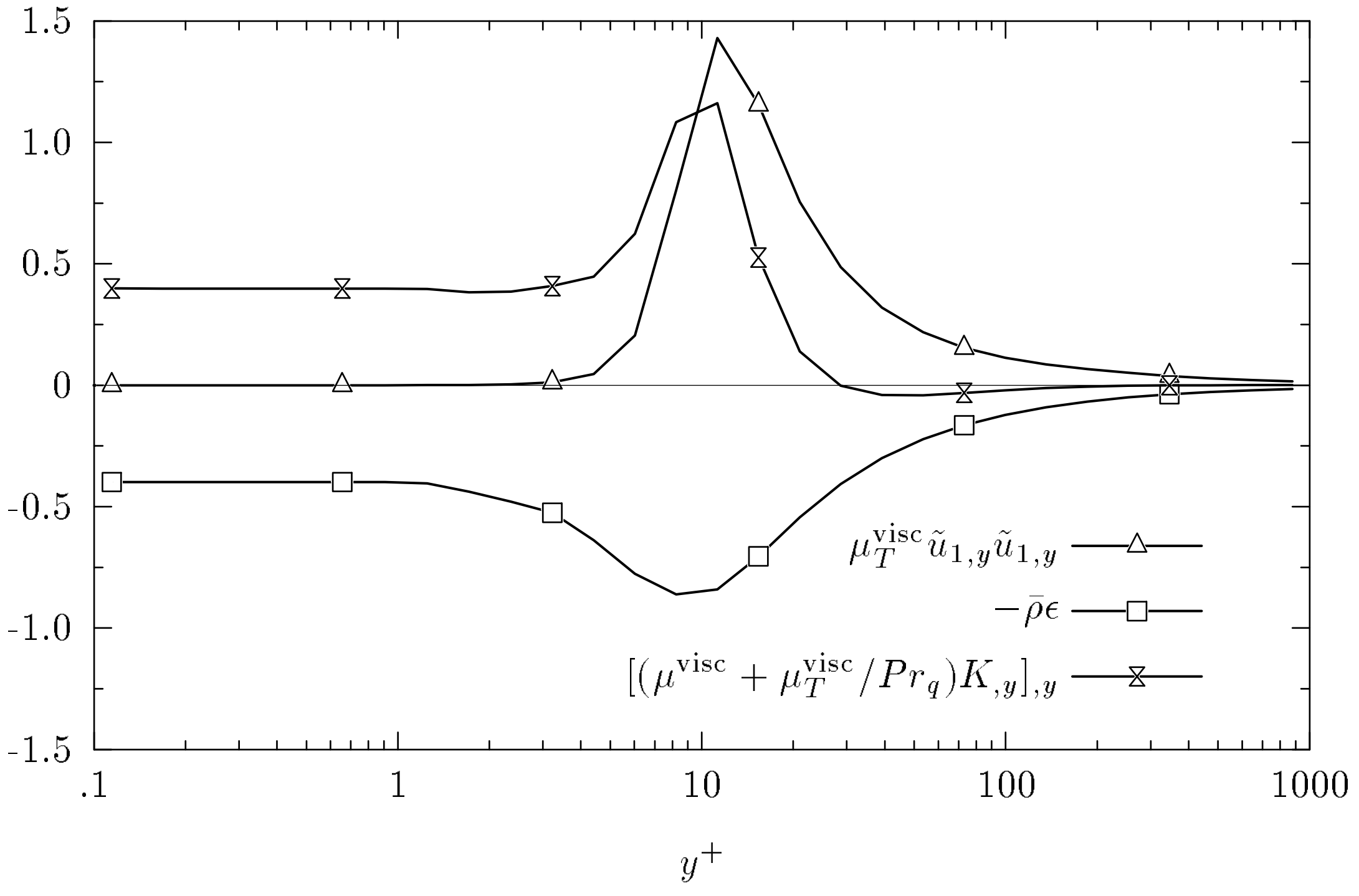

*Figure 8.* Flow over a flat plate. Budget for $k$ at $Re_\theta = 15,144$.

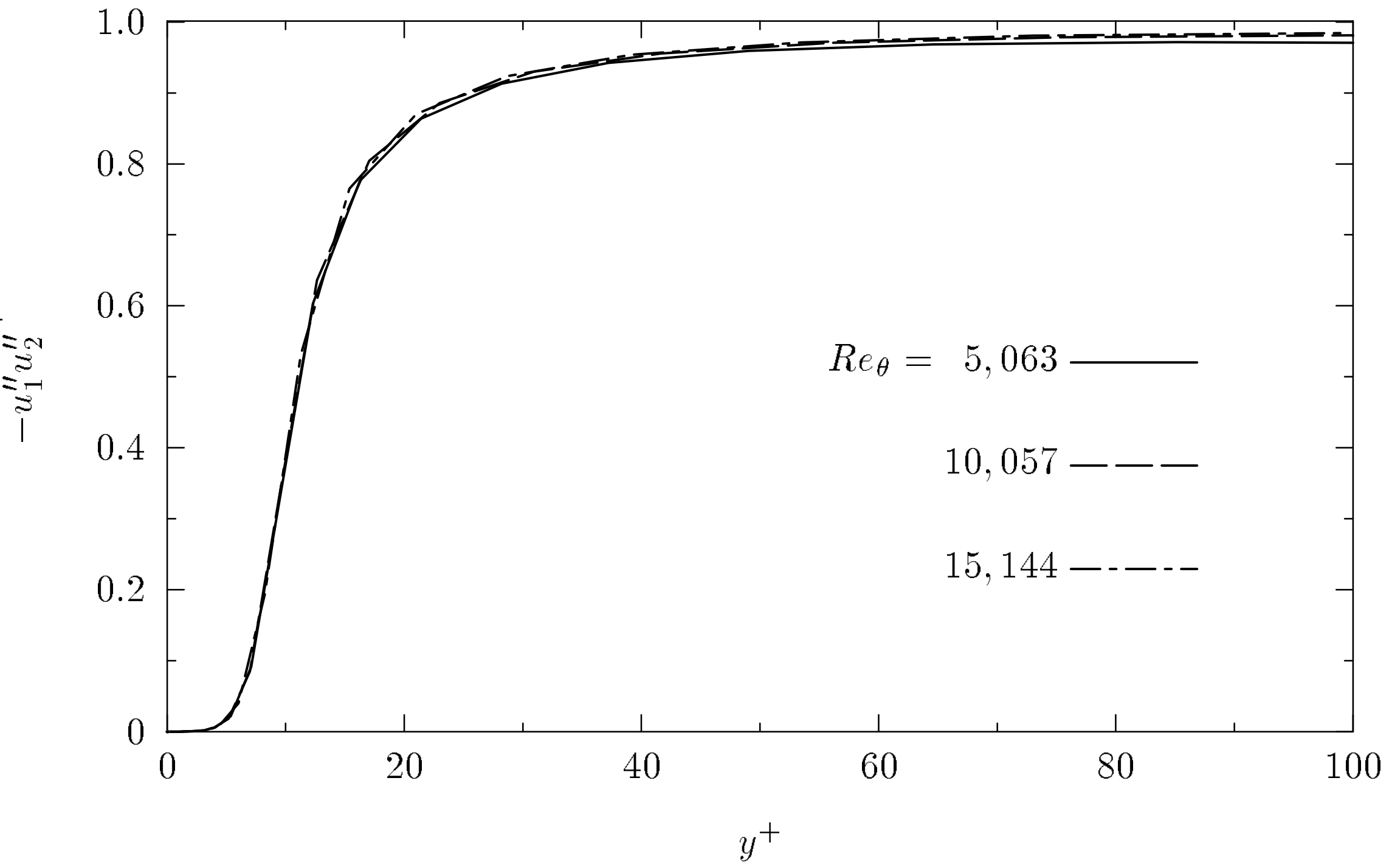

*Figure 9.* Flow over a flat plate. Turbulent stress at several values of $Re_\theta$.

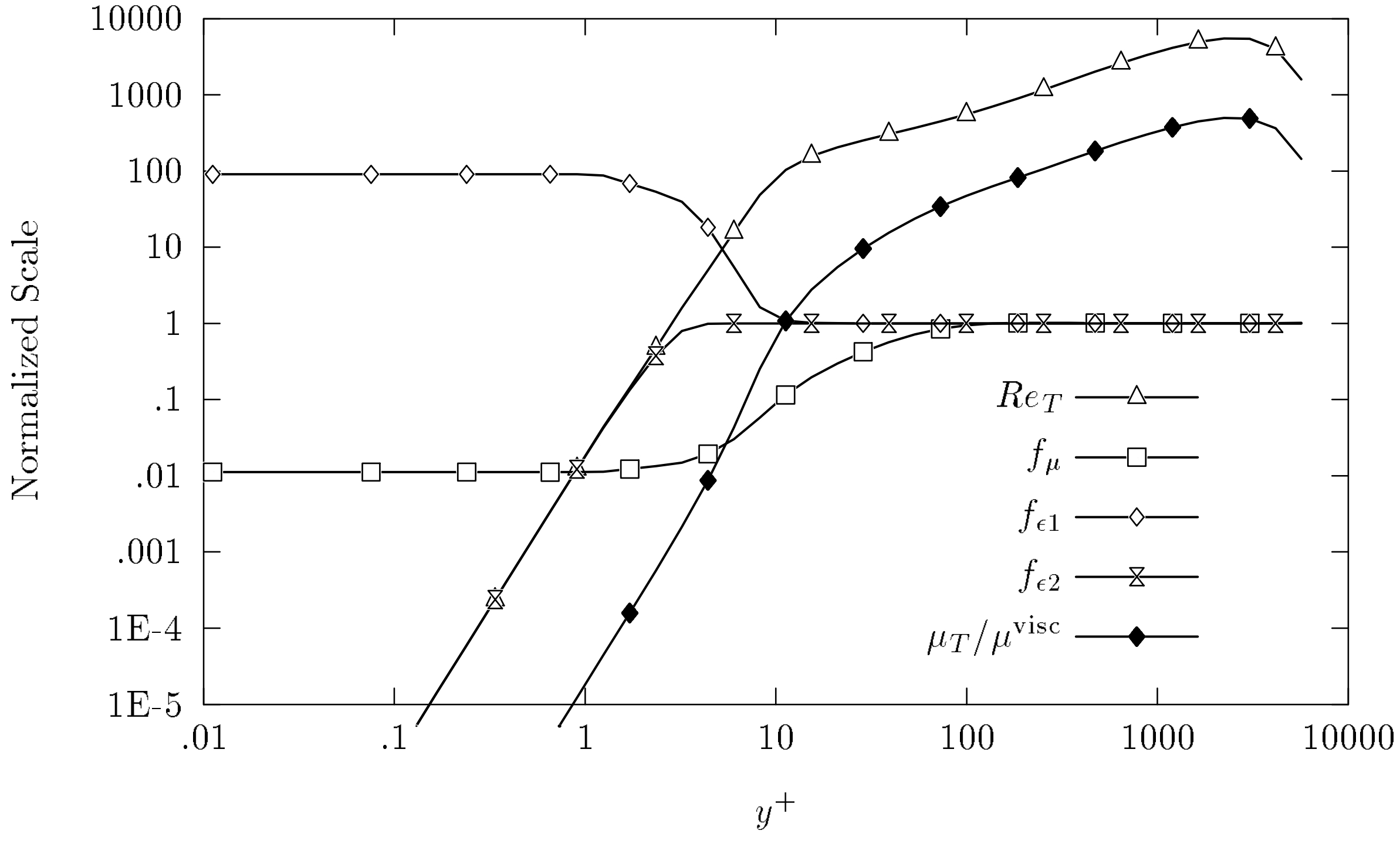

*Figure 10.* Flow over a flat plate. Modeling functions at $Re_\theta = 15,144$.

Variations of temperature have been neglected since the flow is practically isothermal.

### 8.2. Refinement study

A refinement study has been conducted to verify convergence of the solution and to test the impact of element size on accuracy.

Figure 12 shows the influence of the near-wall element size on the friction coefficient. As the size, expressed in the plot in wall coordinates, of the element closest to the wall is increased, the accuracy diminishes, although the effect is small. Even for an element size of $y^+ = 5.0$ the accuracy is reasonable, despite that the source terms for $q_1$ are completely unresolved in the near-wall region. Similarly for the velocity profiles, which are displayed in Figure 13 at a Reynolds number of $Re_\theta = 15,144$. This is unusual for low Re number $k$-$\epsilon$ models, which require a near-wall element size corresponding to less than $y^+ < 1$. This can be attributed to the way the boundary condition for $\epsilon$ is implemented, that is,

$$q_{1\text{wall}}^4 = \nu^2 q_{0,k} q_{0,k} \qquad y^+ < 3. \tag{156}$$

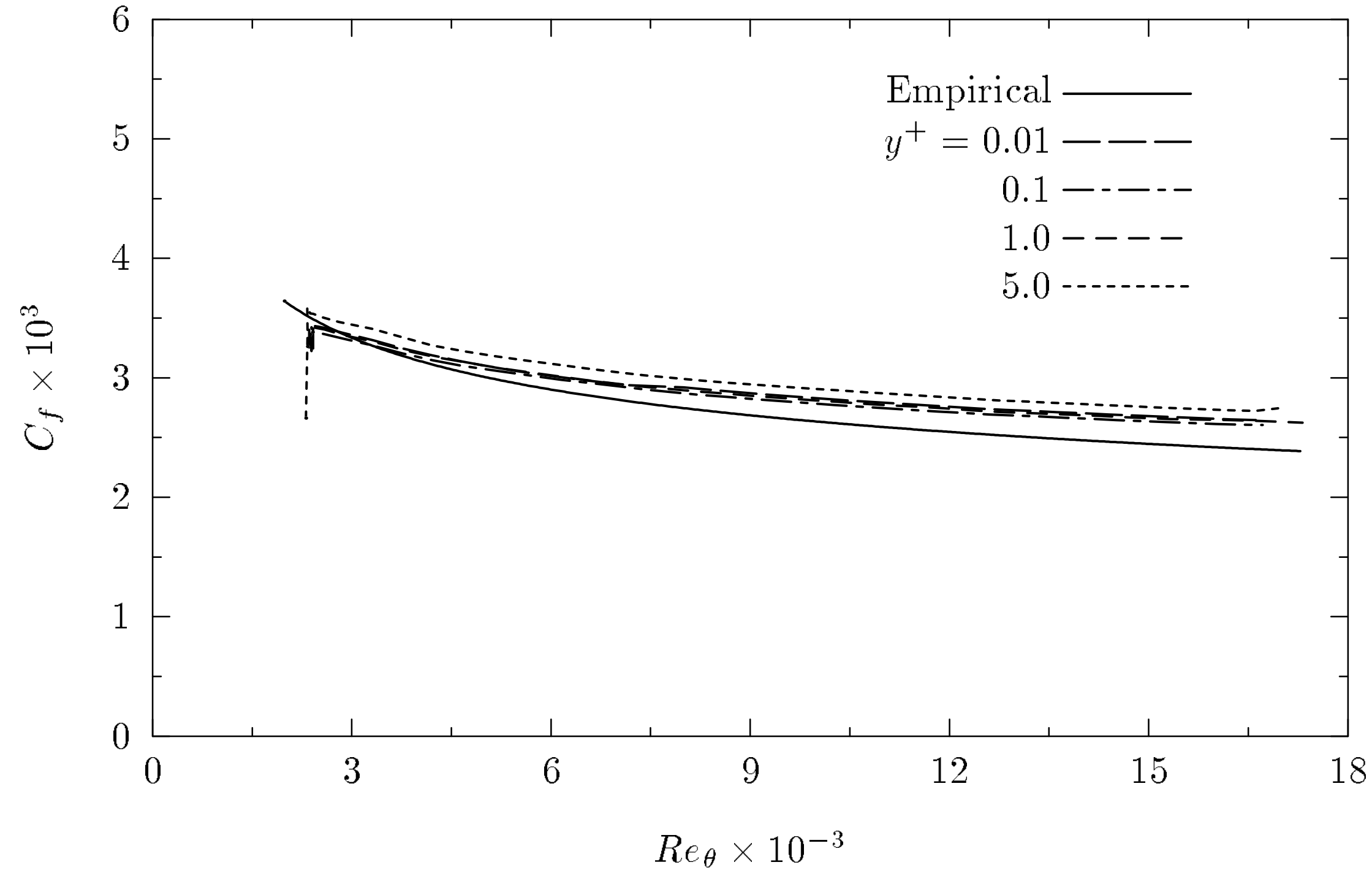

*Figure 12.* Flow over a flat plate. Effect of the location of the nearest node to the wall on the friction coefficient.

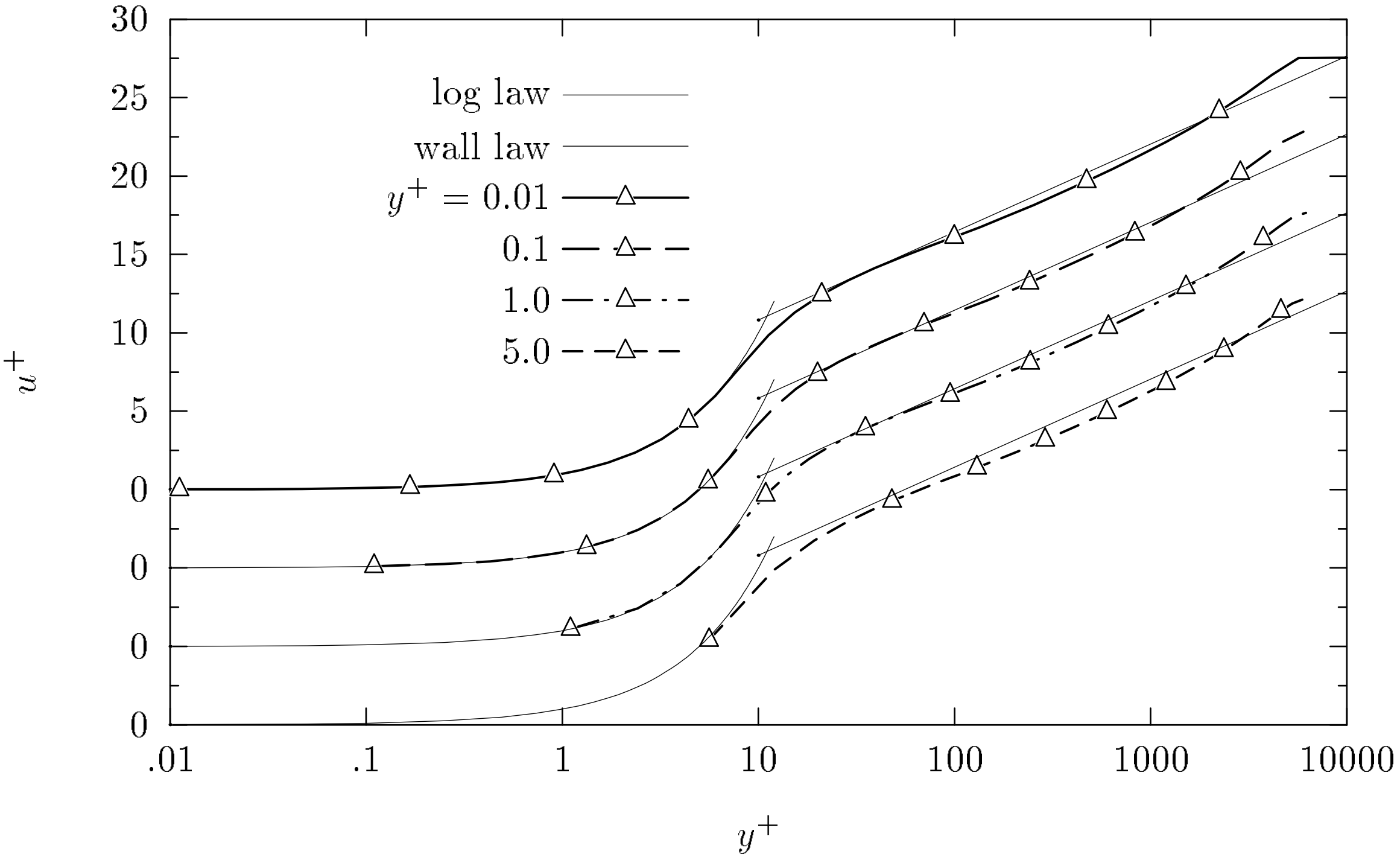

*Figure 13.* Flow over a flat plate. Velocity profiles in wall coordinates for several distances to the wall of the nearest node. $Re_\theta = 15,144$.

## 9. Conclusions

The Reynolds-averaged Navier-Stokes equations coupled with a two-equation turbulence model have been set into the framework of symmetric advective-diffusive systems and the accompanying generalized entropy function has been derived. As a consequence, the system of equations is endowed with an entropy production principle stemming from a generalized second law of thermodynamics. Appropriately defined variational formulations based on this framework inherit at the discrete level an entropy-producing principle, thus providing numerical stability from a nonlinear physical stability principle.

Symmetry and entropy production require that if the hypothesis of turbulent transport of $k$ is respected, i.e. $Pr_{q_1} = Pr_k$, then turbulent diffusivity of $k$ and $\epsilon$ must be equal,

$$Pr_\epsilon = Pr_k$$

a condition that is always employed by some models in the literature and for which positivity of two-equation models require.

The entropy-production principle presented here can also be used to ensure that turbulence modeling assumptions indeed increase the generalized entropy function. Therefore, the developed Clausius-Duhem inequality can be employed to design more physically coherent turbulence models.

The ideas presented here have been tested with the Lam-Bremhorst turbulence model. It has been shown that the model behaved correctly, had little dependency with respect to first-node distance to the wall and that the assumptions introduced lead to good results.

## Acknowledgements

The authors are grateful to Paul Durbin and Jim Stewart for helpful conversations. G. Hauke gratefully acknowledges the financial support through the project

PID2022-138572OB-C44, funded by MCIN/AEI/10.13039/501100011033/FEDER, UE and Gobierno de Aragón/FEDER-UE (Grupo de Referencia de Tecnologias Fluidodinamicas DGA-T32_23R).

## Appendix A. Coefficient matrices for the two-equation turbulence closure

In this Appendix, we consider the system of conservation laws arising when two turbulent velocity scales, defined according to Section 3, are appended to the Reynolds-averaged Navier-Stokes equations. Expressions of the coefficient matrices with respect to the entropy variables and for a general divariant gas are given. Note that tildes and bars have been left out in all expressions for clarity.

For convenience, $\boldsymbol{U}$ and $\boldsymbol{V}$ are repeated here as

$$\boldsymbol{U} = \frac{1}{v}\left\{\begin{array}{c} 1 \\ u_1 \\ u_2 \\ u_3 \\ e + |\boldsymbol{u}|^2/2 + q_0^2/2 + q_1^2/2 \\ q_0 \\ q_1 \end{array}\right\} \tag{A.1}$$

$$\boldsymbol{V} = \frac{1}{T}\left\{\begin{array}{c} \mu - |\boldsymbol{u}|^2/2 - q_0^2/2 - q_1^2/2 \\ u_1 \\ u_2 \\ u_3 \\ -1 \\ q_0 \\ q_1 \end{array}\right\} \tag{A.2}$$

where $\mu$ is the chemical potential. Given $p$, $T$, and $\mu = \mu(p,T)$, we have the following relations:

$$s = -\left(\frac{\partial \mu}{\partial T}\right)_p, \qquad v = \left(\frac{\partial \mu}{\partial p}\right)_T \tag{A.3}$$

$$h = \mu + Ts, \qquad e = h - pv \tag{A.4}$$

$$\alpha_p = \frac{1}{v}\left(\frac{\partial v}{\partial T}\right)_p = \frac{1}{v}\left(\frac{\partial^2 \mu}{\partial p \partial T}\right), \qquad \beta_T = -\frac{1}{v}\left(\frac{\partial v}{\partial p}\right)_T = -\frac{1}{v}\left(\frac{\partial^2 \mu}{\partial p^2}\right)_T \tag{A.5}$$

$$c_p = \left(\frac{\partial h}{\partial T}\right)_p = -T\left(\frac{\partial^2 \mu}{\partial T^2}\right)_p, \qquad c_v = c_p - \frac{\alpha_p^2 v T}{\beta_T} \tag{A.6}$$

We express the coefficient matrices with the help of the following variables:

$$k_{\text{tot}} = \frac{|\boldsymbol{u}|^2 + q_0^2 + q_1^2}{2}, \qquad d = \frac{v\alpha_p T}{\beta_T}, \qquad \bar{\gamma} = \frac{v\alpha_p}{\beta_T c_v}, \tag{A.7}$$

$$c_1 = u_1^2 + \frac{v}{\beta_T}, \qquad c_2 = u_2^2 + \frac{v}{\beta_T}, \qquad c_3 = u_3^2 + \frac{v}{\beta_T}, \tag{A.8}$$

$$c_4 = q_0^2 + \frac{v}{\beta_T}, \qquad c_5 = q_1^2 + \frac{v}{\beta_T}, \tag{A.9}$$

$$\bar{c}_1 = u_1^2 + c_v T, \qquad \bar{c}_2 = u_2^2 + c_v T, \qquad \bar{c}_3 = u_3^2 + c_v T, \tag{A.10}$$

$$\bar{c}_4 = q_0^2 + c_v T, \qquad \bar{c}_5 = q_1^2 + c_v T, \tag{A.11}$$

$$e_1 = h + k_{\text{tot}}, \qquad e_2 = e_1 - d, \qquad e_3 = e_2 + \frac{v}{\beta_T}, \tag{A.12}$$

$$e_4 = e_2 + 2\frac{v}{\beta_T}, \tag{A.13}$$

$$\bar{e}_1 = h - k_{\text{tot}}, \qquad \bar{e}_2 = \bar{e}_1 - d, \qquad \bar{e}_3 = \bar{e}_2 - c_v T, \tag{A.14}$$

$$u_{12} = u_1 u_2, \qquad u_{23} = u_2 u_3, \qquad u_{31} = u_3 u_1, \tag{A.15}$$

$$u_{123} = u_1 u_2 u_3, \qquad a^2 = \frac{v c_p}{c_v \beta_T} \tag{A.16}$$

$$e_5 = e_1^2 - 2e_1 d + \frac{v(2k_{\text{tot}} + c_p T)}{\beta_T}, \qquad \bar{e}_5 = \bar{e}_1^2 - 2\bar{e}_1 d + 2k_{\text{tot}} c_v T + \frac{v c_p T}{\beta_T} \tag{A.17}$$

The Riemannian metric tensor $\boldsymbol{A}_0 = \boldsymbol{U}_{,\boldsymbol{V}}$ and its inverse $\boldsymbol{A}_0^{-1} = \boldsymbol{V}_{,\boldsymbol{U}}$ can be written as

$$\boldsymbol{A}_0 = \frac{\beta_T T}{v^2} \begin{bmatrix} 1 & u_1 & u_2 & u_3 & e_2 & q_0 & q_1 \\ & c_1 & u_{12} & u_{31} & u_1 e_3 & u_1 q_0 & u_1 q_1 \\ & & c_2 & u_{23} & u_2 e_3 & u_2 q_0 & u_2 q_1 \\ & & & c_3 & u_3 e_3 & u_3 q_0 & u_3 q_1 \\ & & & & e_5 & q_0 e_3 & q_1 e_3 \\ & \text{symm.} & & & & c_4 & q_0 q_1 \\ & & & & & & c_5 \end{bmatrix} \tag{A.18}$$

and

$$\boldsymbol{A}_0^{-1} = \frac{v}{c_v T^2} \begin{bmatrix} \bar{e}_5 & u_1 \bar{e}_3 & u_2 \bar{e}_3 & u_3 \bar{e}_3 & -\bar{e}_2 & q_0 \bar{e}_3 & q_1 \bar{e}_3 \\ & \bar{c}_1 & u_{12} & u_{31} & -u_1 & u_1 q_0 & u_1 q_1 \\ & & \bar{c}_2 & u_{23} & -u_2 & u_2 q_0 & u_2 q_1 \\ & & & \bar{c}_3 & -u_3 & u_3 q_0 & u_3 q_1 \\ & & & & 1 & -q_0 & -q_1 \\ & \text{symm.} & & & & \bar{c}_4 & q_0 q_1 \\ & & & & & & \bar{c}_5 \end{bmatrix} \tag{A.19}$$

The advective Jacobian matrices with respect to $\boldsymbol{V}$, $\boldsymbol{A}_i = \boldsymbol{F}^{\text{adv}}_{i,\boldsymbol{V}}$, are given by

$$\boldsymbol{A}_1 = \frac{\beta_T T}{v^2} \begin{bmatrix} u_1 & c_1 & u_{12} & u_{31} & u_1 e_3 & u_1 q_0 & u_1 q_1 \\ & u_1(u_1^2 + 3\frac{v}{\beta_T}) & u_2 c_1 & u_3 c_1 & e_1 \frac{v}{\beta_T} + u_1^2 e_4 & c_1 q_0 & c_1 q_1 \\ & & u_1 c_2 & u_{123} & u_{12} e_4 & u_{12} q_0 & u_{12} q_1 \\ & & & u_1 c_3 & u_{31} e_4 & u_{31} q_0 & u_{31} q_1 \\ & & & & u_1(e_5 + 2e_1 \frac{v}{\beta_T}) & u_1 q_0 e_4 & u_1 q_1 e_4 \\ & \text{symm.} & & & & u_1 c_4 & u_1 q_0 q_1 \\ & & & & & & u_1 c_5 \end{bmatrix} \tag{A.20}$$

$$\boldsymbol{A}_2 = \frac{\beta_T T}{v^2} \begin{bmatrix} u_2 & u_{12} & c_2 & u_{23} & u_2 e_3 & u_2 q_0 & u_2 q_1 \\ & u_2 c_1 & u_1 c_2 & u_{123} & u_{12} e_4 & u_{12} q_0 & u_{12} q_1 \\ & & u_2(u_2^2 + 3\frac{v}{\beta_T}) & u_3 c_2 & e_1 \frac{v}{\beta_T} + u_2^2 e_4 & c_2 q_0 & c_2 q_1 \\ & & & u_2 c_3 & u_{23} e_4 & u_{23} q_0 & u_{23} q_1 \\ & & & & u_2(e_5 + 2e_1 \frac{v}{\beta_T}) & u_2 q_0 e_4 & u_2 q_1 e_4 \\ & \text{symm.} & & & & u_2 c_4 & u_2 q_0 q_1 \\ & & & & & & u_2 c_5 \end{bmatrix} \tag{A.21}$$

$$\boldsymbol{A}_3 = \frac{\beta_T T}{v^2} \begin{bmatrix} u_3 & u_{31} & u_{23} & c_3 & u_3 e_3 & u_3 q_0 & u_3 q_1 \\ & u_3 c_1 & u_{123} & u_1 c_3 & u_{31} e_4 & u_{12} q_0 & u_{12} q_1 \\ & & u_3 c_2 & u_2 c_3 & u_{23} e_4 & u_{23} q_0 & u_{23} q_1 \\ & & & u_3(u_3^2 + 3\frac{v}{\beta_T}) & e_1 \frac{v}{\beta_T} + u_3^2 e_4 & c_3 q_0 & c_3 q_1 \\ & & & & u_3(e_5 + 2e_1 \frac{v}{\beta_T}) & u_3 q_0 e_4 & u_3 q_1 e_4 \\ & \text{symm.} & & & & u_3 c_4 & u_3 q_0 q_1 \\ & & & & & & u_3 c_5 \end{bmatrix} \tag{A.22}$$

We proceed now to the diffusivity matrix $\boldsymbol{K}_{ij}$. The spatial gradients of the velocity components, turbulent velocity scales and temperature are

$$u_{i,j} = T V_{i+1,j} + T u_i V_{5,j} \qquad \text{for } i = 1, 2, 3 \tag{A.23}$$

$$q_{i,j} = T V_{i+6,j} + T q_i V_{5,j} \qquad \text{for } i = 0, 1 \tag{A.24}$$

$$T_{,i} = T^2 V_{5,i} \tag{A.25}$$

Let

$$
\begin{aligned}
\hat{\lambda}^{\text{visc}} &= \lambda^{\text{visc}} - \frac{2}{3}\mu_T^{\text{visc}} \\
\hat{\mu}^{\text{visc}} &= \mu^{\text{visc}} + \mu_T^{\text{visc}} \\
\hat{\kappa} &= \kappa + \kappa_T \\
\mu_k^{\text{visc}} &= \mu^{\text{visc}} + \frac{\mu_T^{\text{visc}}}{Pr_k} \\
\chi^{\text{visc}} &= \hat{\lambda}^{\text{visc}} + 2\hat{\mu}^{\text{visc}}
\end{aligned} \tag{A.26}
$$

The diffusivity coefficient matrices $\boldsymbol{K}_{ij}$, where $\boldsymbol{K}_{ij}\boldsymbol{V}_{,j} = \boldsymbol{F}_i^{\text{diff}}$, are

$$
\boldsymbol{K}_{11} = T \begin{bmatrix}
0 & 0 & 0 & 0 & 0 & 0 & 0 \\
 & \chi^{\text{visc}} & 0 & 0 & \chi^{\text{visc}}u_1 & 0 & 0 \\
 & & \hat{\mu}^{\text{visc}} & 0 & \hat{\mu}^{\text{visc}}u_2 & 0 & 0 \\
 & & & \hat{\mu}^{\text{visc}} & \hat{\mu}^{\text{visc}}u_3 & 0 & 0 \\
 & & & & k_{55}^{11} & \mu_k^{\text{visc}}q_0 & \mu_\epsilon^{\text{visc}}q_1 \\
 & \text{symm.} & & & & \mu_k^{\text{visc}} & 0 \\
 & & & & & & \mu_\epsilon^{\text{visc}}
\end{bmatrix} \tag{A.27}
$$

$$
\boldsymbol{K}_{22} = T \begin{bmatrix}
0 & 0 & 0 & 0 & 0 & 0 & 0 \\
 & \hat{\mu}^{\text{visc}} & 0 & 0 & \hat{\mu}^{\text{visc}}u_1 & 0 & 0 \\
 & & \chi^{\text{visc}} & 0 & \chi^{\text{visc}}u_2 & 0 & 0 \\
 & & & \hat{\mu}^{\text{visc}} & \hat{\mu}^{\text{visc}}u_3 & 0 & 0 \\
 & & & & k_{55}^{22} & \mu_k^{\text{visc}}q_0 & \mu_\epsilon^{\text{visc}}q_1 \\
 & \text{symm.} & & & & \mu_k^{\text{visc}} & 0 \\
 & & & & & & \mu_\epsilon^{\text{visc}}
\end{bmatrix} \tag{A.28}
$$

$$
\boldsymbol{K}_{33} = T \begin{bmatrix}
0 & 0 & 0 & 0 & 0 & 0 & 0 \\
 & \hat{\mu}^{\text{visc}} & 0 & 0 & \hat{\mu}^{\text{visc}}u_1 & 0 & 0 \\
 & & \hat{\mu}^{\text{visc}} & 0 & \hat{\mu}^{\text{visc}}u_2 & 0 & 0 \\
 & & & \chi^{\text{visc}} & \chi^{\text{visc}}u_3 & 0 & 0 \\
 & & & & k_{55}^{33} & \mu_k^{\text{visc}}q_0 & \mu_\epsilon^{\text{visc}}q_1 \\
 & \text{symm.} & & & & \mu_k^{\text{visc}} & 0 \\
 & & & & & & \mu_\epsilon^{\text{visc}}
\end{bmatrix} \tag{A.29}
$$

$$\boldsymbol{K}_{12} = \boldsymbol{K}_{21}^T = T \begin{bmatrix} 0 & 0 & 0 & 0 & 0 & 0 & 0 \\ 0 & 0 & \hat{\lambda}^{\text{visc}} & 0 & \hat{\lambda}^{\text{visc}} u_2 & 0 & 0 \\ 0 & \hat{\mu}^{\text{visc}} & 0 & 0 & \hat{\mu}^{\text{visc}} u_1 & 0 & 0 \\ 0 & 0 & 0 & 0 & 0 & 0 & 0 \\ 0 & \hat{\mu}^{\text{visc}} u_2 & \hat{\lambda}^{\text{visc}} u_1 & 0 & (\hat{\lambda}^{\text{visc}} + \hat{\mu}^{\text{visc}}) u_{12} & 0 & 0 \\ 0 & 0 & 0 & 0 & 0 & 0 & 0 \\ 0 & 0 & 0 & 0 & 0 & 0 & 0 \end{bmatrix} \tag{A.30}$$

$$\boldsymbol{K}_{13} = \boldsymbol{K}_{31}^T = T \begin{bmatrix} 0 & 0 & 0 & 0 & 0 & 0 & 0 \\ 0 & 0 & 0 & \hat{\lambda}^{\text{visc}} & \hat{\lambda}^{\text{visc}} u_3 & 0 & 0 \\ 0 & 0 & 0 & 0 & 0 & 0 & 0 \\ 0 & \hat{\mu}^{\text{visc}} & 0 & 0 & \hat{\mu}^{\text{visc}} u_1 & 0 & 0 \\ 0 & \hat{\mu}^{\text{visc}} u_3 & 0 & \hat{\lambda}^{\text{visc}} u_1 & (\hat{\lambda}^{\text{visc}} + \hat{\mu}^{\text{visc}}) u_{31} & 0 & 0 \\ 0 & 0 & 0 & 0 & 0 & 0 & 0 \\ 0 & 0 & 0 & 0 & 0 & 0 & 0 \end{bmatrix} \tag{A.31}$$

$$\boldsymbol{K}_{23} = \boldsymbol{K}_{32}^T = T \begin{bmatrix} 0 & 0 & 0 & 0 & 0 & 0 & 0 \\ 0 & 0 & 0 & 0 & 0 & 0 & 0 \\ 0 & 0 & 0 & \hat{\lambda}^{\text{visc}} & \hat{\lambda}^{\text{visc}} u_3 & 0 & 0 \\ 0 & 0 & \hat{\mu}^{\text{visc}} & 0 & \hat{\mu}^{\text{visc}} u_2 & 0 & 0 \\ 0 & 0 & \hat{\mu}^{\text{visc}} u_3 & \hat{\lambda}^{\text{visc}} u_2 & (\hat{\lambda}^{\text{visc}} + \hat{\mu}^{\text{visc}}) u_{23} & 0 & 0 \\ 0 & 0 & 0 & 0 & 0 & 0 & 0 \\ 0 & 0 & 0 & 0 & 0 & 0 & 0 \end{bmatrix} \tag{A.32}$$

where

$$\begin{aligned} k_{55}^{11} &= \chi^{\text{visc}} u_1^2 + \hat{\mu}^{\text{visc}} (u_2^2 + u_3^2) + \hat{\kappa} T + \mu_k^{\text{visc}} q_0^2 + \mu_\epsilon^{\text{visc}} q_1^2 \\ k_{55}^{22} &= \chi^{\text{visc}} u_2^2 + \hat{\mu}^{\text{visc}} (u_1^2 + u_3^2) + \hat{\kappa} T + \mu_k^{\text{visc}} q_0^2 + \mu_\epsilon^{\text{visc}} q_1^2 \\ k_{55}^{33} &= \chi^{\text{visc}} u_3^2 + \hat{\mu}^{\text{visc}} (u_2^2 + u_1^2) + \hat{\kappa} T + \mu_k^{\text{visc}} q_0^2 + \mu_\epsilon^{\text{visc}} q_1^2 \end{aligned} \tag{A.33}$$